\documentclass[a4paper,11pt]{article}
\pdfoutput=1 
\usepackage{jheppub} 
\usepackage[T1]{fontenc} 
\usepackage{mathrsfs}
\newcommand{\AdS} {{AdS}}
\usepackage{shuffle}
\usepackage{musicography}
\usepackage{empheq}
\usepackage{mathrsfs}
\usepackage{bm}
\usepackage{verbatim}
 \usepackage{makecell}
 \usepackage{float}
 \usepackage{hyperref}
 \usepackage[makeroom]{cancel}

\newcommand{\be}{\begin{equation}}
\newcommand{\ee}{\end{equation}}
\newcommand{\bpm}{\begin{pmatrix}}
\newcommand{\epm}{\end{pmatrix}}

\newcommand{\EV}[1]{\langle #1 \rangle}
\newcommand{\beqn}{\begin{eqnarray}}
\newcommand{\eeqn}{\end{eqnarray}}

\usepackage{slashed}

\DeclareMathOperator{\Res}{Res}

\usepackage{pgffor}
\foreach \x in {A,...,Z}{
  \expandafter\xdef\csname b\x\endcsname{\noexpand\mathbb{\x}}
}
\foreach \x in {A,...,Z}{
  \expandafter\xdef\csname c\x\endcsname{\noexpand\mathcal{\x}}
}
\foreach \x in {A,...,Z}{
  \expandafter\xdef\csname s\x\endcsname{\noexpand\mathscr{\x}}
}
\foreach \x in {A,...,Z}{
  \expandafter\xdef\csname sf\x\endcsname{\noexpand\mathsf{\x}}
}
\foreach \x in {a,...,z}{
  \expandafter\xdef\csname sf\x\endcsname{\noexpand\mathsf{\x}}
}
\foreach \x in {A,...,Z}{
  \expandafter\xdef\csname  fk\x\endcsname{\noexpand\frak{\x}}
}
\foreach \x in {a,...,z}{
  \expandafter\xdef\csname  fk\x\endcsname{\noexpand\frak{\x}}
}
\newcommand{\wt}{\widetilde}

 \newcommand{\bv}{ \begin{verbatim}}

\newcommand{\p}{\partial}

\newcommand{\ii}{\mathrm{i}}
\newcommand{\dd}{\mathrm{d}}

\title{D1-D5 CFT data from $AdS_3 \times S^3$ Virasoro-Shapiro amplitude  }

\abstract{
The AdS Virasoro--Shapiro amplitude has recently been generalized to the   AdS$_3$/CFT$_2$ correspondence between type IIB string theory on $\AdS_3 \times S^3 \times K3$ (or $T^4$), supported by Ramond--Ramond flux, and the D1--D5 CFT. 
In this paper, we use the $\AdS\times S$ Virasoro--Shapiro machinery to extract strong-coupling CFT data of the D1--D5 CFT by extending and completing earlier analyses in several directions. 
First, starting from the superconformal/Mellin block expansion of four-point functions of half-BPS tensor operators with arbitrary external KK modes, we employ the full $\AdS\times S$ Mellin formalism to bootstrap the $\AdS_3 \times S^3$ Virasoro--Shapiro amplitude for general KK configurations. 
This establishes its consistency with superconformal symmetry and yields a wealth of additional CFT data naturally organized in internal Mellin space. 
Second, we push the computation to the next order in the strong-coupling expansion and extract additional higher-order CFT data. 
Third, we translate the resulting Mellin-space data into the internal spin basis. 
We derive the transformation kernel relating internal Mellin variables and $SU(2)_L \times SU(2)_R$ R-symmetry spins.
 As applications, we obtain explicit formulae for the scaling dimensions of long multiplets on the first two leading Regge trajectories  
 of arbitrary internal spins,
and certain three-point   functions  with   half-BPS tensor operators.
These results provide a valuable set of analytic D1--D5 CFT data, enabling future applications and direct comparison  with complementary approaches such as integrability.
 }

\author[a,b ]{Hongliang Jiang} 
\affiliation[a]{Center for Mathematics and Interdisciplinary Sciences, Fudan University, Shanghai 200433, China }
\affiliation[b]{Shanghai Institute for Mathematics and Interdisciplinary Sciences (SIMIS), Shanghai 200433, China} 

\emailAdd{jianghongliang@fudan.edu.cn }

\begin{document}
\maketitle
 
\section{Introduction and summary}
  AdS$_3$/CFT$_2$ duality is a fascinating correspondence between three-dimensional quantum gravity and two-dimensional conformal field theory. It has a long history and continues to be a very active area of research. The origins of the correspondence can be traced back to the seminal work  \cite{Brown:1986nw}    forty years ago, where the emergence of Virasoro symmetry and the associated central charge were derived purely from the perspective of three-dimensional gravity. Remarkably, this insight predates by more than a decade the celebrated proposal of the AdS/CFT correspondence   \cite{Maldacena:1997re}.

Beyond its historical significance, AdS$_3$/CFT$_2$ duality is an exceptionally rich subject. On the gravity side, three-dimensional quantum gravity can be studied quantitatively using effective field theory methods---both in the Einstein--Hilbert formulation and in its Chern--Simons description---as well as within the framework of string theory via a worldsheet approach. On the field theory side, the infinite-dimensional Virasoro symmetry provides powerful additional leverage for understanding two-dimensional conformal field theory. These special features are largely absent in higher-dimensional holographic dualities, making AdS$_3$/CFT$_2$ a uniquely tractable arena for probing the delicate and genuinely quantum aspects of gravity.

  In recent years, there has been remarkable progress toward deriving the AdS$_3$/CFT$_2$ duality directly from string theory. In particular, the precise correspondence between type IIB string theory on $\AdS_3 \times S^3 \times M_4$ with minimal NS--NS flux and the symmetric orbifold CFT $\mathrm{Sym}^N(M_4)$ has been established from several complementary perspectives \cite{Eberhardt:2018ouy,Eberhardt:2019ywk}. This progress relies crucially on the existence of a WZW model description of the worldsheet theory for strings on $\AdS_3$ supported purely by NS--NS flux.
However, the restriction to pure NS--NS flux substantially limits the scope of this approach. Extending these methods to a broader class of AdS$_3$/CFT$_2$ dualities---in particular to backgrounds supported by RR flux, or more generally by mixed NS--NS/RR flux---remains a major challenge.~\footnote{For mixed flux,   one can start with pure NS--NS flux and then incorporate the effects of RR flux perturbatively.  This gives a very indirect way to generalize to the mixed flux case in principle. }   
Consequently, a central open problem in establishing AdS$_3$/CFT$_2$ duality in full generality is to understand backgrounds supported by RR flux. In this case, the absence of an equally powerful and tractable worldsheet formulation makes a direct derivation appear out of reach.

Nevertheless, an indirect yet powerful approach to studying string theory in the presence of RR flux has been proposed. 
Rather than constructing a microscopic worldsheet formulation of such a theory, one adopts the so-called  {single-valued assumption} 
for the worldsheet representation of certain observables. 
This method was first applied to $\AdS_5 \times S^5$ in the study of graviton scattering \cite{Alday:2022uxp,Alday:2022xwz,Alday:2023jdk,Alday:2023mvu}, 
where an ansatz was proposed for the relevant worldsheet integrals in terms of  {single-valued multiple polylogarithms} (SVMPLs). 
The unknown coefficients in this ansatz were then uniquely fixed by imposing various consistency conditions, 
leading to an AdS analogue of the flat-space Virasoro--Shapiro amplitude. 
The resulting object is therefore referred to as the  {AdS Virasoro--Shapiro (VS) amplitude}. 
This framework also provides a wealth of CFT data for $\cN=4$ super-Yang--Mills theory.

  This method was subsequently generalized to the study of the $\AdS_3/\mathrm{CFT}_2$ correspondence in \cite{Chester:2024wnb,Jiang:2025oar}. 
  \footnote{See also \cite{Alday:2023pzu,Chester:2024esn,Alday:2024rjs,Wang:2025pjo,Ren:2026zxs} for other applications and generalizations.
 }
The goal of this paper is to extend these results   in several directions, with the aim of extracting new CFT data for two-dimensional conformal field theories. 
Let us begin by reviewing the basic setup. 
We consider type IIB string theory on $\AdS_3\times S^3\times M_4$, where $M_4=K3$ or $T^4$. 
This background arises as the near-horizon limit of the D1--D5 system: one wraps $Q_1$ D1-branes along a non-compact direction, and $Q_5$ D5-branes along $M_4$ together with the same non-compact direction shared with the D1-branes. 
The resulting $\AdS_3$ geometry is supported by Ramond--Ramond (RR) flux.
The holographic dual is a two-dimensional $\mathcal N=(4,4)$ superconformal field theory (SCFT) describing the low-energy dynamics of the D1--D5 bound state, commonly referred to as the D1--D5 CFT. 
The curvature scale of AdS and the   coupling in the CFT are related by~\footnote{See \cite{Aharony:2024fid,David:2002wn} for more details about the D1-D5   system.   }
\begin{equation}
\frac{L_{\text{AdS}}^2}{L_{\text{string}}^2}
= g\sqrt{Q_1 Q_5}
\equiv \sqrt{\lambda} ~,
\qquad\qquad
\alpha' = L_{\text{string}}^2~,
\end{equation}
where $L_{\text{AdS}}$ is the $\AdS_3$ radius, $L_{\text{string}}$ is the string length, and $g$ is the six-dimensional string coupling. 
Thus, the 't~Hooft-like limit corresponds to $\lambda\to\infty$, or equivalently $L_{\text{AdS}}/L_{\text{string}}\to\infty$. 
In this regime, the strong-coupling expansion of the CFT matches the curvature (i.e.\ $\alpha'$) expansion on the gravity side.
Although this setup has been known for a long time, dating back to \cite{Maldacena:1997re}, the corresponding D1--D5 CFT remains rather mysterious.  
In particular, it admits  no simple general description---for instance, in terms of a Lagrangian formulation---except at certain special points on its moduli space.  This stands in sharp contrast to the $\AdS_5/\mathrm{CFT}_4$ correspondence, where the dual field theory is simply $\cN=4$ super-Yang--Mills theory. This raises a basic and important question: what are the defining CFT data of the elusive D1--D5 CFT?

The AdS Virasoro--Shapiro machinery comes to the rescue. 
In \cite{Chester:2024wnb}, by studying the lowest-lying half-BPS tensor operator, the worldsheet representation of the AdS Virasoro--Shapiro amplitude was derived, together with a wealth of CFT data. 
This was subsequently generalized to half-BPS tensor operators in the full Kaluza--Klein (KK) tower in \cite{Jiang:2025oar}, made possible by the Mellin-space techniques developed in \cite{Wang:2025pjo}. 
However, several questions remain open. 
First, the AdS Virasoro--Shapiro amplitude in \cite{Jiang:2025oar} was initially derived in the special case of $\EV{pp11}$, and then extended to general KK modes using crossing symmetry in Mellin space. 
It is therefore not immediately obvious whether the resulting amplitude for arbitrary KK external states is fully consistent with the underlying superconformal symmetry, or more concretely with the superconformal block expansion. 
Second, while \cite{Jiang:2025oar} extracted a class of CFT data, this analysis was largely restricted to the $\EV{pp11}$ correlator, leaving the general KK configurations unexplored.

The central goal of this paper is to clarify these issues and to extract as much CFT data as possible, in a form amenable to comparison with other approaches. 
We study the superconformal (Mellin) block expansion for general configurations of external half-BPS tensor operators, i.e.\ for arbitrary KK modes. 
This enables us to rederive the $\AdS\times S$ Virasoro--Shapiro amplitude by imposing the single-valued ansatz together with a set of consistency conditions. 
\footnote{This approach was also used by \cite{Wang:2025owf} in the study of gluon scattering. }
Our derivation reproduces the result of \cite{Jiang:2025oar}, thereby confirming its compatibility with the superconformal block expansion, while simultaneously yielding a broader set of CFT data relevant to generic four-point functions of KK tensor modes.

We also push the analysis to the next order in the strong-coupling expansion. 
On the string worldsheet side, the would-be contribution at this order vanishes;   enforcing this vanishing turns into a nontrivial consistency requirement and allows us to extract additional higher-order CFT data. 

Finally, since the CFT data obtained from the $\AdS\times S$ Virasoro--Shapiro machinery are naturally produced in Mellin space, we translate them into the internal spin basis, obtaining results as functions of the $SU(2)_L\times SU(2)_R$ R-symmetry spins. 
We derive the transformation kernel relating the internal Mellin variables to the internal spin variables, outline a systematic procedure for carrying out this conversion, and illustrate it with simple examples. 
In particular, we find that the anomalous dimension of a  long multiplet on the first two leading Regge trajectories is given universally by  \eqref{tau2jjb} and \eqref{tau2jjbsb}.
As another application, we present  the closed form expressions in \eqref{pptauOPE} for the  three-point OPE coefficients between two identical half-BPS tensor operators and a long multiplet on the leading Regge trajectory.
These results provide concrete targets for future comparisons with other approaches, such as integrability.
 
 Let us close with a few future directions. 
It would be interesting to push the computation to higher orders in the curvature expansion. 
At the next nontrivial order, the worldsheet integrals become genuinely more involved and require weight-six SVMPLs \cite{Alday:2022xwz}, which should in turn allow access to additional CFT data. 
It would also be important to generalize the machinery to other half-BPS operators in the gravity multiplet, whose kinematic and superconformal structures are considerably more involved. 
The supergravity limit for such operators has recently been studied in \cite{Aprile:2025kfk}.

The rest of the paper is organized as follows. 
In Section~\ref{AdS4pt}, we introduce our setup from the $\AdS_3$ string theory perspective and review the kinematic structure of four-point functions in two-dimensional SCFTs, both in position space and in Mellin space, as well as their Borel transforms. 
In Section~\ref{AdSVS}, we analyze the general block expansion and derive the $\AdS \times S$ Virasoro--Shapiro amplitude for generic correlators $\EV{p_1p_2p_3p_4}$. 
We also extract the corresponding CFT data in Mellin space. 
In Section~\ref{MelSpin}, we translate the CFT data from the internal Mellin variables to the internal spin basis. 
Finally, a number of technical details and useful formulas are collected in the appendices.

\section{Superconformal kinematics} \label{AdS4pt}

\subsection{Tensor multiplet in $\AdS_3\times S^3$}
 We consider type IIB string theory on \( \AdS_3 \times S^3 \times M_4 \), where \( M_4 = K3 \). The low-energy limit is described by six-dimensional \( \mathcal{N} = (2,0) \) supergravity on \( \AdS_3 \times S^3 \), coupled to \( h_{1,1} + 1 \) tensor multiplets, with \( h_{1,1}(K3) = 20 \).  \footnote{We focus on the \( K3 \) case in the main text. In the case of $M_4=T^4$, we have $\cN=(2,2)$ supergravity and  \( h_{1,1}(T^4) = 4 \).}
Further compactifying on $S^3$ gives rise to various towers of KK modes in $\AdS_3$. See \cite{Rastelli:2019gtj} for a summary.
 
The symmetry relevant for our discussion    is given by $PSU(1,1|2)_L\times PSU(1,1|2)_R$, corresponding to the left- and right-moving sector. Its bosonic subgroup, $ SL(2,\mathbb R)_L\times SU(2)_L\times SL(2,\mathbb R)_R\times SU(2)_R $,   matches the isometry group $SO(2,2)\times SO(4) $ of the $\AdS_3 \times S^3$ background.  
Each operator is labeled by the quantum numbers $h,j,\bar h, \bar j$, corresponding to the charges under the Cartan generators of the bosonic subgroup. The scaling dimension $\Delta$ and spacetime spin $\ell$ of the operator are then given by $\Delta=h+\bar h, \ell=h-\bar h$, while $j,\bar j$ are the internal spins  under the R-symmetry  $SU(2)_L\times SU(2)_R $.
As in higher-dimensional CFTs, we also define the twist of an operator   as $\tau=\Delta-\ell$. 

In this paper, we are interested in the $1/2$-BPS operators coupling to the tensor multiplets in the supergravity, whose superconformal primaries are denoted as $s_p^I$, where $p=1, 2,\cdots$ labels the KK modes under the reduction on $S^3$, and $I=1, \cdots, h_{1,1}+1$ labels the vector representation of the flavor symmetry group $SO(h_{1,1}+1)$. The quantum numbers are given by $h=j=\bar h =\bar j=p/2$.

 The flavor symmetry $SO(h_{1,1}+1)$ is valid in the supergravity limit and is respected by the cubic coupling. However, in full string theory, this symmetry is broken down to $SO(h_{1,1})$. See e.g. \cite{Taylor:2007hs}.   We will therefore focus on the unique $SO(h_{1,1})$ singlet, which we refer to as the tensor operator and denote by $\mathcal O_p$.
 For $p=1$, the superconformal descendants of  $\cO_1$ include a marginal operator that couples to the string coupling, which is dual to the dilaton of string theory. In the planar limit that we consider, this   multiplet   does not care about the compact spacetime  $M_4$   and is invariant under   certain  duality symmetries, including T-duality  \cite{Chester:2024wnb}. For higher $p>1$, $\cO_p$ can be regarded as the cousins    of $\cO_1$ with higher KK modes under the reduction on $S^3$.  
 
\subsection{Position space}
We will discuss the four-point functions of tensor operators in position space and superconformal constraints in this subsection. 

The operator $\cO_p$ transforms as spin-$(p/2,p/2)$ representation under the $SU(2)_L \times SU(2)_R$ R-symmetry, and   carries $p$ symmetric vector indices under $SO(4)$. To simplify notation, we introduce an index-free form
\be
\cO_p(\sfx;\sfy) \coloneq \cO_p^{\mu_1 \cdots \mu_p} (\sfx) \sfy_{\mu_1}  \cdots \sfy_{\mu_p}~,
\ee
where $\sfy _\mu$ is an $SO(4)$ null vector satisfying $\sfy_\mu \sfy^\mu=0$. Using the isomorphism $SO(4)\simeq SU(2) \times SU(2)$, we can parametrize $\sfy^\mu=\sigma_{\alpha\dot\alpha}^\mu y^\alpha \bar y ^{\dot \alpha}$ and label $\cO$ in spinor representation 
 \be
 \cO_p(\sfx; \sfy )=  \cO_p(\sfx; y,\bar y ) =\cO_p^{\alpha_1\cdots\alpha_p, \dot\alpha_1\cdots \dot\alpha_p}
 y_{\alpha_1}\cdots y_{\alpha_p}\;\bar y_{\dot\alpha_1}\cdots \bar y_{\dot\alpha_p}~.
 \ee
 Similarly, instead  of using the real Euclidean coordinate $\sfx^1, \sfx^2$ to label the position of operators in the dual 2d CFT, we introduce complex coordinates $z=\sfx^1+ \ii\, \sfx^2,\bar z=\sfx^1-\ii\, \sfx^2$. The Lorentz-invariant distance between two points is then given by
 \be
 \sfx_{ij}^2=(\sfx^1_i-\sfx_j^1)^2+(\sfx^2_i-\sfx_j^2)^2=z_{ij}\bar z_{ij}~, \quad \text{with} \quad z_{ij}=z_i -z_j~, \;\bar z_{ij}=\bar z_i -\bar z_j\,~.
 \ee  
For the R-symmetry polarization vectors, we define analogously $\sfy_{ij}=\sfy_i^\mu \sfy _{ j\,  \mu}= y_{ij}  \bar y_{ij}$, where $y_{ij}=\epsilon_{\alpha\beta} y_i^\alpha y_j^\beta, \bar y_{ij}=\epsilon_{\dot\alpha\dot\beta}  \bar y_i^{\dot\alpha}\bar   y_j^ {\dot \beta}$. 

We can construct conformal and R-symmetry cross ratios from the kinematic building blocks introduced above. There are two conformal cross ratios, $U$ and $V$, and two R-symmetry cross ratios, $\sigma$ and $\rho$, defined as
\be
U=\frac{\sfx_{12}^2 \sfx_{34}^2}{\sfx_{13}^2 \sfx_{24}^2}=z \bar{z}~, 
\qquad V=\frac{\sfx_{14}^2 \sfx_{23}^2}{\sfx_{13}^2 \sfx_{24}^2}=(1-z)(1-\bar{z})~ , 
\ee
and 
\be\label{yybrel}
 \sigma =\frac{\sfy_{13} \sfy_{24}}{\sfy_{12} \sfy_{34}}=\frac{1}{y \bar y} ~, 
\qquad
 \rho =\frac{\sfy_{14} \sfy_{23}}{\sfy_{12} \sfy_{34}}=(1-1/y)(1-1/\bar{y})~ . 
 \ee
In terms of spinor variables, the cross ratios are equivalently given by:
\be
z=\frac{z_{12} z_{34}}{z_{13} z_{24}}~, \qquad \bar{z}=\frac{\bar{z}_{12} \bar{z}_{34}}{\bar{z}_{13} \bar{z}_{24}}~, \qquad  \frac{1}{y}= \frac{y_{13} y_{24}}{y_{12} y_{34}}~, \qquad  \frac{1}{\bar y}=\frac{\bar{y}_{13} \bar{y}_{24}}{\bar{y}_{12} \bar{y}_{34}}~ .
\ee

Finally, in superconformal kinematics, it is useful to define the combinations:
\be\label{gijdef}
g_{ij}=\frac {y_{ij}}{z_{ij}}~, \qquad \bar g_{ij}=\frac{\bar y_{ij}}{\bar z_{ij}} ~, \qquad
 \bm g_{ij}=g_{ij} \bar g_{ij}=\frac {\sfy_{ij}}{\sfx_{ij}^2}~.
\ee

We would like to study the four-point function which admits the following factorization structures: 
\be\label{cOKG}
\EV{ \cO_{p_1}(\sfx_1; \sfy_1 )  \cO_{p_2}(\sfx_2; \sfy_2 )
  \cO_{p_3}(\sfx_3; \sfy_3 )   \cO_{p_4}(\sfx_4; \sfy_4 ) }
 = {\bf K }  \cG (z, \bar z ; y, \bar y)
 = \wt{\bf K } \wt \cG (z, \bar z ; y, \bar y)~,
\ee
which will be abbreviated as $\EV{p_1p_2p_3p_4}$. 
Here we introduce two types of factorization forms which are convenient for different purposes. 

The prefactors  $\mathbf{K}$, $\wt{\mathbf{K}}$ capture  the kinematic dependence and are given by

   \beqn\label{KKtd}
\wt{\bf K}&=&\bm g_{12}^{\frac{p_1+p_2 }{2}}\bm g_{34}^{\frac{p_3+p_4 }{2}} 
\Big(\frac{\bm g_{14}}{ \bm g_{24}}\Big)^{\frac{p_1-p_2}{2}}
\Big(\frac{\bm g_{14}}{ \bm g_{13}}\Big)^{\frac{p_4-p_3}{2}}
=\bm g_{12}^{\frac{p_1+p_2 }{2}}\bm g_{34}^{\frac{p_3+p_4 }{2}} 
\Big(\frac { \bm g_{24}} {\bm g_{14}}\Big)^{\frac{p_ {21}}{2}}
\Big(\frac { \bm g_{13}}{\bm g_{14}}\Big)^{\frac{p_{34}}{2}}
\\
\label{KK}
{\bf K}&=&\bm g_{12}^{\frac{p_1+p_2-p_3-p_4}{2}}\bm g_{13}^{\frac{p_1+p_3-p_2-p_4}{2}}\bm g_{14}^{p_4}\bm \;\bm g_{23}^{\frac{p_2+p_3+p_4-p_1}{2}}
\Big(\frac{\bm g_{12}\bm g_{34}} {\bm g_{14}\bm g_{23}}\Big) ^L
\\&=&  \label{Kfactor}
\bm g_{12}^{\frac{p_1+p_2-p_3-p_4}{2}}\bm g_{13}^{\frac{p_1+p_3-p_2-p_4}{2}} \bm \;\bm g_{23}^{\frac{p_2+p_3-p_4-p_1}{2}}(\bm g_{14}\bm g_{23})^{p_4}
\Big(\frac{V}{U \rho } \Big) ^L~ .
\eeqn
 where $p_{ij} =p_i-p_j$ and $L$ is called  extremality and is defined as
\be\label{extermL}
L=\min_i(p_i,\Sigma -p_i)~, \qquad
\Sigma = \frac12 (p_{1}+p_{2}+p_{3}+p_{4})  ~.
 \ee
 Note that for a physically legitimate four-point correlator, we should have $L\ge 1$. In addition,  note  that $\Sigma$ must be an integer.

  
The two factors $\bf K$  and $\wt {\bf K}$   are related in a simple way: 
  \beqn\label{KKtdrel}
\wt{\bf K}&=&  {\bf K } \Big(\frac{\bm g_{12}\bm g_{34}}{\bm g_{14}\bm g_{23}} \Big) ^{\frac{p_3+p_4-2L}{2}}
 \Big(\frac{\bm g_{14}\bm g_{23}}{\bm g_{13}\bm g_{24}} \Big) ^{\frac{p_1-p_2}{2}}
={\bf K } \Big(\frac{ V}{U\rho} \Big) ^{\frac{p_3+p_4-2L}{2}}
 \Big(\frac{\rho }{V\sigma} \Big) ^{\frac{p_1-p_2}{2}}~.
\eeqn

 For later purposes, we will also introduce  the following   notation to separate the  contributions from    position space and internal space
 \be\label{KKxy}
{ \bf K}=K_\sfy K_\sfx~, \qquad \wt{ \bf K}=\wt K_\sfy\wt K_\sfx~,
 \ee
 where $K_\sfy, \wt K_\sfy$ only depends on   $\sfy_{ij}$ or equivalently $y_{ij}, \bar y_{ij}$, while $K_\sfx,\wt K_\sfx$ only depends on $\sfx^2_{ij}$ or  equivalently $z_{ij}, \bar z_{ij}$.

 The function $\cG$  in \eqref{cOKG} satisfies the superconformal Ward identities and admits the decomposition \cite{Rastelli:2019gtj}:
\be\label{cgs}
\cG=\cG_0 +(1-z/y) (1-\bar z/ \bar y)  \cH(U,V; \sigma, \rho)~,
\ee
where $\cG_0$ is the protected part that obeys
$
\cG_0(z,\bar z; y, \bar y=\bar z)=f(z,y),  
\cG_0(z,\bar z; y= z, \bar y )=f(\bar z,\bar y) ,
$
and survives in the free limit due to non-renormalization theorems.

\subsection{Mellin space}
 
 We now switch to Mellin space.
Following \cite{Aprile:2025kfk},  we first reparametrize the position space correlator as follows:
\be\label{cOdecomp} 
 \EV{ \cO_{p_1}(\sfx_1; \sfy_1 )  \cO_{p_2}(\sfx_2; \sfy_2 )
  \cO_{p_3}(\sfx_3; \sfy_3 )   \cO_{p_4}(\sfx_4; \sfy_4 ) }
 =\text{protected}+(z-y)(\bar z-\bar y)  \sfx_{13}^2 \sfx_{24}^2 \sfy_{13}^2 \sfy_{24}^2 \mathscr H~.\quad
\ee
Then  we consider the following $AdS \times S$ Mellin transformation
\be\label{HMellin}
 \mathscr H (\sfx_{ij},\sfy_{ij})=\int  \Big( \prod_{i<j}\frac{\dd\delta_{ij}}{(2\pi i)^2}  \Big) 
\sum_{m_{ij}\in\bZ}\prod_{i<j} \frac{\Gamma(\delta_{ij})}{\Gamma(m_{ij}) }  \sfx_{ij}^{-2\delta_{ij}} \sfy_{ij}^{ m_{ij}}
\cM(\delta_{ij},m_{ij})~.
\ee
The integration and summation variables are subject to some constraints:
\be\label{detaij}
\sum_i \delta_{ij} =0~, \qquad
\delta_{ij}=\delta_{ji}~, \qquad
\delta_{ii}=-p_i-1~,
\ee
\be\label{mijfs}
\sum_i m_{ij} =0~, \qquad
m_{ij}=m_{ji}~, \qquad
m_{ii}=1- p_i ~.
\ee

The integration   over $\delta_{ij}$ is the well-known Mellin transform for CFT correlator in position space \cite{Mack:2009mi}, while the summation   over $m_{ij}$ is the analogue in internal R-symmetry space \cite{Aprile:2020luw}.
\bigskip
\subsubsection*{Hidden conformal symmetry}
\bigskip

 Before moving to the details, let us pause by  mentioning one nice property underlying  \eqref{HMellin}. If we assume that the Mellin amplitude has the following property:
\be\label{Mthetadm}
\cM_{p_1p_2p_3p_4}(\delta_{ij},m_{ij})=\cM(\theta_{ij} ) ~, \qquad
\theta_{ij}=\delta_{ij}-m_{ij}~.
\ee
Then we can easily see that
\beqn
\mathscr H (\sfx_{ij}^2,\sfy_{ij})
&=&\int \dd\delta_{ij}\sum_{m_{i j}}
\cM_{p_1p_2p_3p_4}(\delta_{ij},m_{ij})
\prod_{i<j}   
\frac{\Gamma(\delta_{ij}) }{\Gamma(m_{ij}+1)}
\frac{\sfy_{ij}^{m_{ij}}}{   (\sfx_{ij}^2)^{ \delta_{ij}}}
\\ &=&
 \sum_{m_{i j}}
\int \dd\theta_{ij}  
\cM (\theta_{ij} ) 
\prod_{i<j}    
\frac{\Gamma(\theta_{ij}+m_{ij}) }{\Gamma(m_{ij}+1)}
\frac{\sfy_{ij}^{m_{ij}}}{   (\sfx_{ij}^2)^{ \theta_{ij}+m_{ij}} }~.
 \eeqn
Note that the summation  over $m_{ij}$ is constrained due to \eqref{mijfs}. This constraint can be relaxed by summing over all $p_i$, which allows us to sum over all $m_{ij} $ independently.  
More explicitly, we have
 \beqn
 \widehat  {  \mathscr H}(\sfx_{ij}^2,\sfy_{ij})=
\sum_{p_i}\mathscr H_{p_1p_2p_3p_4}(\sfx_{ij}^2,\sfy_{ij})
  &=&
  \int \dd\theta_{ij} 
\cM(\theta_{ij} ) 
\prod_{i<j}     \sum_{m_{ij}}
\frac{\Gamma(\theta_{ij}+m_{ij}) }{\Gamma(m_{ij}+1)}
\frac{\sfy_{ij}^{m_{ij}}}{   (\sfx_{ij}^2)^{ \theta_{ij}+m_{ij}} }
\nonumber
\\&=&
\int \dd\theta_{ij} 
\cM(\theta_{ij} ) 
\prod_{i<j}   
\frac{\Gamma(\theta_{ij} ) } {   (\sfx_{ij}^2-\sfy_{ij})^{ \theta_{ij} } }~,
 \eeqn
where we used the identity followed from Taylor expansion:
 \be\sum_{ m\in \bZ}
 \frac{\Gamma(\theta+m)}{\Gamma(m+1)}\frac{ \sfy^m}{(\sfx^2)^{\theta+m}}=\frac{\Gamma(\theta)}{(\sfx^2-\sfy)^\theta}~.
 \ee

Therefore, we see that $   \widehat  {  \mathscr H}$ only depends on the 6d distance $\sfx_{ij}^2-\sfy_{ij}$, namely
$   \widehat  {  \mathscr H}(\sfx_{ij}^2,\sfy_{ij})=   \widehat  {  \mathscr H}(\sfx_{ij}^2-\sfy_{ij})$.
This implies the correlator has  a certain kind of hidden conformal symmetry.

\bigskip
 \subsubsection*{Reduction to independent variables}
 \bigskip
 
Since the integration and summation variables in \eqref{cOdecomp} are not independent, we would like to simplify the previous $AdS\times S$  Mellin transform. 

\bigskip 

\noindent{\bf AdS part.} Let us first simplify the $\AdS$ part or external Mellin part.
From \eqref{detaij}, we see that there are 6 variables subject to 4 constraints. So there are just 2 independent variables, which can be chosen freely. It turns out to be very convenient to choose the    $s,t$ defined below as independent variables:
\beqn
s&=& p_1+p_2-2\delta_{12}= p_3+p_4-2\delta_{34}=\Sigma-\delta_{12}-\delta_{34}~,
\\
t&=&  p_1+p_4-2\delta_{14}= p_2+p_3-2\delta_{23}=\Sigma-\delta_{14}-\delta_{23}~, 
\\
 \tilde u&=&  p_1+p_3-2\delta_{13 }= p_2+p_4-2\delta_{24}=\Sigma-\delta_{13}-\delta_{24}~. 
 \eeqn
 Note that 
 \be
 s+t+
 \tilde u=2\Sigma -2~.
 \ee
We can now simplify the Mellin transformation in \eqref{HMellin}. For the position space part, one can show that (we will ignore the integrand and mainly focus on the measure below)
\beqn
&& \int \frac{\dd s\, \dd t}{(2\pi i)^2} 
\prod_{i<j}  {\Gamma(\delta_{ij})}{  }  \sfx_{ij}^{-2\delta_{ij}}  
\\&=&
\frac{1}{\sfx_{13}^2 \sfx_{24}^2}
(\sfx_{12}^2)^{-\frac{  p_{1 }+p_2}{2}} ( \sfx_{34}^2)^{-\frac{  p_{3 }+p_4}{2}}
\Big( \frac{\sfx_{24}^2}{\sfx_{14}^2}\Big) ^{\frac{p_1-p_2}{2}}  
\Big( \frac{\sfx_{14}^2}{\sfx_{13}^2}\Big) ^{\frac{p_3-p_4}{2}}  
\int  \frac{\dd s\, \dd t}{(2\pi i)^2}
U^{\frac{s }{2}}V^{\frac{t-p_2-p_{3}}{2}}  
 \nonumber  \\ & &
  \times  \Gamma\Big(\frac{p_{12}-s}{2} \Big)  \Gamma\Big(\frac{p_{34}-s}{2} \Big)  \Gamma\Big(\frac{p_{14}-t}{2} \Big)  \Gamma\Big(\frac{p_{23}-t}{2} \Big)  \Gamma\Big(\frac{p_{13}-\tilde u}{2} \Big)  \Gamma\Big(\frac{p_{24}-\tilde u}{2} \Big)
\nonumber
\\&=&
\frac{1}{\sfx_{13}^2 \sfx_{24}^2}
\wt K_\sfx
\int  \frac{\dd s\, \dd t}{(2\pi i)^2}
U^{\frac{s }{2}}V^{\frac{t-p_2-p_{3}}{2}}   \Gamma_\AdS (s,t)~,
 \label{MellinAdS}
\eeqn
where $\wt K_\sfx$ is defined via \eqref{gijdef},   \eqref{KKtd} and \eqref{KKxy}, and
\beqn\label{GmAdS}
\Gamma_{AdS}(s,t)&=& \Gamma\Big(\frac{p_1+p_2-s}{2} \Big)  \Gamma\Big(\frac{p_3+p_4-s}{2} \Big) 
\Gamma\Big(\frac{p_1+p_4-t}{2} \Big)  \Gamma\Big(\frac{p_2+p_3-t}{2} \Big)
\\ &&
\times
 \Gamma\Big(\frac{p_1+p_3-\tilde u}{2} \Big)  \Gamma\Big(\frac{p_2+p_4-\tilde u}{2} \Big)~ . 
\eeqn

\bigskip

 \noindent{\bf S part.} Next, we consider the $S$ part or internal Mellin space part. 
From \eqref{mijfs},   we similarly have 2 free parameters, which can be chosen as $m_{24},m_{34}$. The rest are then given by 
\begin{align}\label{m123422}
m_{12} &= \frac12\left(p_{1} + p_{2} - p_{3} - p_{4} + 2 m_{34}\right)~ ,\\
m_{13} &= \frac12\left(p_{1} - p_{2} + p_{3} - p_{4} + 2 m_{24}\right)~ ,\\
m_{14} &= -1 + p_{4} - m_{24} - m_{34}~ ,\\
m_{23} &= \frac12\left(-2 - p_{1} + p_{2} + p_{3} + p_{4} - 2 m_{24} - 2 m_{34}\right)~ .
\end{align}

Just like the $s,t$   introduced before, it is also more convenient to consider
  \be\label{mstu}
   m_s=m_{12}+m_{34}~,  \quad
   m_t=m_{14}+m_{23}~,  \quad
   m_u=m_{13}+m_{24}~. \quad
   \ee  
   which are subject to the condition 
   \be
   m_s+m_t+m_u
=\Sigma-2~.
   \ee
   Then  all the $m_{ij}$ can be expressed in terms of $m_s,m_t$ and $p_i$:
   \beqn\label{m1234}
   m_{24}&=\frac12 \left(-2-m_s -m_t+ p_2 +p_4 \right) ~, 
    \qquad
      m_{34}&=\frac14 \left(2m_s -p_1-p_2+p_3+p_4 \right) ~,
\\
m_{12}&=\frac{1}{4}\left(2m_s+p_1+p_2-p_3-p_4\right)~,\qquad
m_{13}&=\frac{1}{2}\left(-2-m_s-m_t+p_1+p_3\right)~,
\\
m_{14}&=\frac{1}{4}\left(2m_t+p_1-p_2-p_3+p_4\right)~,\qquad
m_{23}&=\frac{1}{4}\left(2m_t-p_1+p_2+p_3-p_4\right)~.\qquad
\eeqn
Note that  $m_s,m_t$ are not just integers; instead they are even integers shifted by 0 or 1:
\be
m_s\in 2\bZ+\frac12(p_1+p_2-p_3-p_4)\subset \bZ~,\qquad
m_t\in 2\bZ+\frac12(-p_1+p_2+p_3-p_4)\subset \bZ~,\qquad
\ee
which can be seen from the condition $m_{ij}\in \bZ$ and $\Sigma\in \bZ$.  We will use both 
 $(m_{24},m_{34})$ and $(m_s,m_t)$ as independent variables interchangeably, whenever convenient.

For the internal space, the Mellin transformation simplifies to
\beqn
&&\sum_{m_{ij}\in\bZ}\prod_{i<j} \frac{\sfy_{ij}^{m_{ij}} }{\Gamma(m_{ij}+1) } 
\\&=&\sum_{m_{ij}}  
\frac{
\left(\frac{\sfy_{12}\sfy_{34}}{\sfy_{14}\sfy_{23}}\right)^{m_{34}}
\left(\frac{\sfy_{13}\sfy_{24}}{\sfy_{14}\sfy_{23}}\right)^{m_{24}}
\sfy_{13}^{\frac12(p_1-p_2+p_3-p_4)}
\sfy_{12}^{\frac12(p_1+p_2-p_3-p_4)}
\sfy_{23}^{\frac12(-p_1+p_2+p_3-p_4)}
(\sfy_{14}\sfy_{23})^{p_4-1}
}{
\prod_{i<j}\Gamma(m_{ ij}+1)
}
\nonumber\\[6pt]
&=&\sum_{m_{ij}} 
\frac{
\sigma^{m_{24}}\rho^{-m_{24}-m_{34}}\,
\sfy_{12}^{\frac12(p_1+p_2-p_3-p_4)}
\sfy_{13}^{\frac12(p_1-p_2+p_3-p_4)}
\sfy_{23}^{\frac12(-p_1+p_2+p_3-p_4)}
(\sfy_{14}\sfy_{23})^{p_4-1}
}{
\prod_{i<j}\Gamma(m_{ ij}+1)
}
\nonumber\\[6pt]
&=&\frac{\rho}{\sfy_{14}\sfy_{23}}
\sum_{m_{ij}} 
\frac{
\sigma^{m_{24}}\rho^{L-1-m_{24}-m_{34}}\, 
}{
\prod_{i<j}\Gamma(m_{ ij}+1)
} 
\sfy_{12}^{\frac12(p_1+p_2-p_3-p_4)}
\sfy_{13}^{\frac12(p_1-p_2+p_3-p_4)}
\sfy_{23}^{\frac12(-p_1+p_2+p_3-p_4)}
(\sfy_{14}\sfy_{23})^{p_4}/\rho^{L}
\nonumber\\[6pt]
\\&=&
\frac{K_\sfy}{\sfy_{12}\sfy_{34}}
\sum_{m_{24},m_{34} } 
\frac{ \,\sigma^{m_{24}}\rho^{L-1-m_{24}-m_{34}}}
{\Gamma_{S}(m_s,m_t)}~,
\label{MellinSS}
\eeqn
where  $K_\sfy$ is given in \eqref{KK} and \eqref{KKxy} and 
\beqn\label{GmS}
\Gamma_{S}(m_s,m_t)&=&
\prod_{i<j}\Gamma(m_{ ij}+1)
\\&=&  \Gamma \left( \frac{m_s}{2} - \frac{\gamma}{2} + 1 \right) 
   \Gamma \left( \frac{m_s}{2} + \frac{\gamma}{2} + 1 \right) 
   \Gamma \left( \frac{m_t}{2} + \frac{\alpha}{4} - \frac{\beta}{4} + 1 \right)  
    \Gamma \left( \frac{m_t}{2} + \frac{\beta}{4} - \frac{\alpha}{4} + 1 \right) 
\nonumber\\& &\times   \Gamma \left(  \frac{m_u}{2}  - \frac{\alpha}{4} - \frac{\beta}{4} +1 \right) 
    \Gamma \left(  \frac{m_u}{2}  + \frac{\alpha}{4} + \frac{\beta}{4} +1 \right) ~.
\eeqn
Although the summation is over all integers, it actually truncates to a finite sum due to   the Gamma function in the denominator, which requires all $m_{ij}\ge 0$ \eqref{m123422}. In particular, $m_{14},m_{23}\ge 0$ which implies that 
\be
m_{24}+m_{34} \le  \min(p_4, \frac{p_2+p_3+p_4 -p_1}{2})-1~.
\ee
 
If we order the $p_i$ such that  $p_1\ge \{ p_2 , p_3 \}\ge p_4$
\footnote{Here  $p_1\ge \{ p_2 , p_3 \}\ge p_4$ is equivalent to  $p_1\ge p_2\ge p_4$ and
$p_1\ge p_3\ge p_4$, while the ordering of $p_2,p_3$ does not matter. 
This should be contrasted with the ordering $p_1\ge p_2$ and $p_3\ge p_4$, which is more general. 
}
, then this is exactly $m_{24}+m_{34} \le L-1$, where the extremality $L$ is given in \eqref{extermL}.
So effectively the summation is $m_{24},m_{34}\in \bN, m_{24}+m_{34} \le L-1$.
This particularly implies that \eqref{MellinSS} is a polynomial of $\rho,\sigma$ with degree  $L-1$.

Combining the Mellin transformation in both $\AdS$  \eqref{MellinAdS} and $S$ part \eqref{MellinSS}, we get
\beqn\label{cHsst}
 \mathscr H &=&\int \frac{\dd s \dd t}{(2\pi i)^2} 
\sum_{m_{ij}\in\bZ}\prod_{i<j} \frac{\Gamma(\delta_{ij})}{\Gamma(m_{ij}+1) }  \sfx_{ij}^{-2\delta_{ij}} \sfy_{ij}^{ m_{ij}}
\cM
\\&=&
  \frac{{\bf K}}{x_{13}^2 x_{24}^2\sfy_{12}\sfy_{34}}
\int  \frac{ds\,dt}{(2\pi i)^2}  \sum_{m_{24},m_{34} } 
U^{\frac{s-p_3-p_4 }{2}+L}V^{\frac{t-p_1+p_{4}}{2}-L} 
\frac{   \Gamma_\AdS(s,t)  }
{\Gamma_{S}(m_s,m_t)}
\sigma^{m_{24}}\rho^{L-1-m_{24}-m_{34}}
\cM~,
 \label{cHsst2}
\qquad\qquad
\eeqn
where we used
\be
\wt K_\sfx K_\sfy=
K_\sfx K_\sfy \Big(\frac{ V}{U } \Big) ^{\frac{p_3+p_4-2L}{2}}
 \Big(\frac{1 }{V } \Big) ^{\frac{p_1-p_2}{2}}
 ={\bf K} U^{L-\frac{p_3+p_4  }{2}}V^{-L+\frac{-p_1+p_2+p_3+p_4 }{2}}~,
\ee

The overall coefficient in \eqref{cHsst2} can be combined with coefficients in \eqref{cOdecomp} to give
\be
(z-y)(\bar z-\bar y)\sfx_{13}^2\sfx_{24}^2 \sfy_{13}  \sfy_{24}   \times   \frac{{1}}{\sfx_{13}^2 \sfx_{24}^2\sfy_{12}\sfy_{34}}
=(z-y)(\bar z-\bar y)\frac{\sfy_{13}  \sfy_{24} }{\sfy_{12}\sfy_{34}}
=\frac{(z-y)(\bar z-\bar y)}{y\bar y}~. \quad
\ee

By comparing  \eqref{cHsst2} with   \eqref{cOKG} \eqref{cgs},  we see that 
\be
 \cH =
 \int  \frac{\dd s\,\dd t}{(2\pi i)^2}  \sum_{m_{24},m_{34} } 
U^{\frac{s-p_3-p_4 }{2}+L}V^{\frac{t-p_1+p_{4}}{2}-L} 
\frac{   \Gamma_ {AdS}(s,t)  }
{\Gamma_{S}(m_s,m_t)}
\sigma^{m_{24}}\rho^{L-1-m_{24}-m_{34}}
\cM(s,t,m_s, m_t )~.
\ee
 
It turns out to be also useful to separate the two types of Mellin transformations into two steps, and consider
\beqn \label{MellinExt}
\hspace{-8ex} 
\cH(U,V,\sigma, \rho )&=&\int \frac{\dd s\, \dd t}{(2\pi \ii)^2}\, U^{\frac{s-p_3-p_4}{2}+L}\, V^{\frac{t+p_4-p_1}{2}-L }\,    
\Gamma_{AdS}(s,t)\; M(s,t,\sigma, \rho )~,
\\
 M(s,t,\sigma, \rho )&=&  \sum_{m_{24},m_{34} } 
 \frac{  \sigma^{m_{24}}\rho^{L-1-m_{24}-m_{34}}  }
{\Gamma_{S}(m_s,m_t)}
 \cM(s,t,m_s, m_t )~. \label{MellinInt}
 \eeqn

 \subsection{Borel transform}
 To really get an AdS analogue of flat space scattering amplitude, one also needs to perform the so-called Borel transform \cite{Penedones:2010ue}. 
 The Borel transformation in our case  is given by 
\be \label{AST}
\cA(S,T, m_s, m_t )
=
\lambda^{\tfrac12}\,
\Gamma(\Sigma)
\int \frac{d\eta}{2\pi i}\,
\frac{e^{\eta}}{\eta^{\Sigma+1}}\,
\cM\!\left(
\frac{2\sqrt{\lambda}\, S}{\eta}
+
\frac{2\Sigma - 2}{3},
\;
\frac{2\sqrt{\lambda}\, T}{\eta}
+
\frac{2\Sigma - 2}{3},
\;
 m_s, m_t 
\right),
\ee
where  $S+T+U=0$. In addition, to simplify some expressions below, we introduce  the following set of variables:
\be\label{abgp}
\alpha=p_1-p_2, \quad \beta=p_3-p_4, \quad \gamma=\frac12(p_1+p_2-p_3-p_4), \quad
\Sigma = \frac12 (p_{1}+p_{2}+p_{3}+p_{4})   ~.\qquad
\ee

\section{$AdS  \times S $ Virasoro-Shapiro amplitude} \label{AdSVS}
In this section, we will make full use of the $\AdS\times S$ Mellin formalism to bootstrap the $\AdS_3\times S^3$ Virasoro--Shapiro amplitude for arbitrary external KK configurations and to extract the associated CFT data in Mellin space.

 \subsection{Flat limit }
 Let us first look at some structures of $\cA$ in \eqref{AST}. In particular, it admits   strong coupling expansion or curvature expansion:
 \be\label{Aexpansion}
 \cA(S,T,m_s, m_t ) 
=\sum_{i=0}^{\infty} \frac{\cA^{(i)}(S,T,m_s, m_t)}{\lambda^{i/2}}~, \qquad \quad 
  \sqrt{\lambda} = \frac{L_{\text{AdS}}^2}{L_{\text{string}}^2}  ~.
\ee
Note that  $\lambda$ only appears  in the form of various powers of $\sqrt{\lambda}$, as   one can see from  \eqref{AST}. 
This in particular implies that 
\be\label{Afracpw}
\cA^{(\frac12)}=\cA^{(\frac32)}=\cdots=0~.
\ee
 
Meanwhile,   the  leading term in \eqref{Aexpansion} should correspond to the flat space  Virasoro-Shapiro amplitude which is computable from the flat space string theory via the worldsheet integral~\footnote{Here the measure is defined as $\dd^2z= \dd z \dd \bar z /(-2\pi \ii)$.} \cite{Jiang:2025oar}
\beqn   \label{flatVS} 
\cA^{(0)}(S, T) 
&= &
\int \,{\dd^2 z  \; |z|^{-2S-2}}{|1 - z|^{-2T-2}}
\frac{S^2+T^2+U^2}{12U^2} 
+ (S \leftrightarrow T)+ (S \leftrightarrow U )~
\\&=&
-\left(S^2+T^2+U^2\right)\frac{\Gamma (-S) \Gamma (-T) \Gamma (-U)  }{4 \Gamma (S+1) \Gamma (T+1) \Gamma (U+1)}~,
\label{A0STVS}
\eeqn
where $S+T+U=0$. 

 \subsection{Supergravity limit}
 
In the supergravity limit, the $\AdS \times S$ Mellin amplitude takes a remarkably simple form \cite{Wen:2021lio} 
\be \label{sugraM2}
  \cM_{\text{SG}}(s,t;m_ {s},m_t)=-\frac{1}{\bm s+2}-\frac{1}{\bm t+2}-\frac{1}{\bm u+2}~,
  \ee
 where the boldfaced Mandelstam variables are defined as:
   \beqn \label{boldstu}
   \bm s &=& s-p_1-p_2+2 m_{12}=s-p_3-p_4+2 m_{34}=s-   \Sigma  +m_s~ ,
\\
   \bm t &=& t-p_1-p_4+2 m_{14}=t-p_2-p_3+2 m_{23}
   =t-   \Sigma  +m_t~,
\\
   \bm u &=& \tilde  u-p_1-p_3+2 m_{13}=\tilde  u-p_2-p_4+2 m_{24}
   =\tilde  u-    \Sigma  +m_u\, ,
   \eeqn
 which are subject to:
   \be
   \bm s+\bm t+\bm u=-4~ .
   \ee 
 
As a result,  \eqref{sugraM2} has exactly the structure in \eqref{Mthetadm}, which signals  the hidden conformal symmetry in the supergravity limit.

 We now apply the Borel transform \eqref{AST}  to the supergravity part \eqref{sugraM2}, and get the  following  supergravity contribution  
\be\label{A1SG}
\cA_{\text{SG}} = \frac{S^2+T^2+U^2}{ 4 S T U }
-\frac{  {  \Sigma } -1 }{12\sqrt{\lambda }}
\Big[     \frac{\Sigma-3   {m_s}-4}{S^2}+
\frac{\Sigma-3   {m_t} -4}{T^2}
+
\frac{\Sigma-3 m_u -4}{U^2}
\Big]+ \cO(1/\lambda)~.
\ee

\subsection{Mellin block and curvature expansion}

 Given the superconformal block expansion of correlation function in position space, we can perform the Mellin transform and obtain the Mellin block expansion of the Mellin amplitude.  The superconformal block expansion will be described in section~\ref{superbkexp}. In particular, we can apply the   Mellin transform \eqref{MellinExt} to the spacetime part the superconformal block in \eqref{EUV}, and get the  the following Mellin block 
\be \label{melblock}
\cQ_s(w;\tau, \ell, m)=\kappa_{\ell,m,\tau+2, d=2}^{(p_1+1,p_2+1,p_3+1,p_4+1)}
Q_{\ell,m}^{p_1-p_2,p_3-p_4,\tau+2}(w-p_1-p_4 )~,
\ee
where  the expression $Q_{\ell,m}^{\Delta_{12},\Delta_{34},\tau}(s)$ and $\kappa_{\ell,m,\tau,d}^{(p_1,p_2,p_3,p_4)}$ are given in Appendix~\ref{apd-mackpol}.

 Then the Mellin amplitude can be expanded in terms of the Mellin block as follows: 
 
\be\label{Mst}
\cM(s,t,m_s,m_t)
=
\sum_{\tau ,\ell}
\cC_{s}(\tau ,\ell,m_s,m_t)
\sum_{m=0}^{\infty}
\frac{
\cQ_{s}(\tilde{u};\, \tau ,\, \ell,\, m)
}{
s - \tau  - 2m
}
\;+\;\cdots, \qquad   
\ee
where $\tilde u =p_{1}+p_{2}+p_{3}+p_{4}-2-s-t$ and the dots represent the rest of  contribution which are not singular in the $S$-channel. The summation above is performed  over all exchanged multiplets in the $S$-channel with twist $\tau$ and spacetime spin $\ell$. Since we consider the   four-point correlator/amplitude with arbitrary scaling dimensions, it suffices to focus on the $S$-channel, and the rest of $T$, $U$-channels can be easily obtained from the permutation of the external operators.\footnote{For example, the $T$-channel  result can be obtained from that of $S$-channel by exchanging $p_2\leftrightarrow p_4,S \leftrightarrow T$.} Furthermore,  the coefficient  $C_{s}(\tau ,\ell,m_s,m_t)$ encodes the OPE coefficients from various exchanged operators and particularly depend on internal space variables $m_s,m_t$.  

Substituting \eqref{Mst} into   \eqref{AST} to perform the Borel transformation, we get the AdS scattering amplitude
 
\be
\cA(S,T,m_s, m_t)=
\sum_{\tau ,\ell}
\cC_{s}(\tau ,\ell,m_s, m_t)A_{\tau,\ell}(S,T )|_{S\text{-pole}}+\cdots
\ee
where
\beqn
A_{\tau,\ell}(S,T )|_{S\text{-pole}}
&=& 
\lambda^{\tfrac12}\,
\Gamma(\Sigma)
\int \frac{d\eta}{2\pi i}\,
\frac{e^{\eta}}{\eta^{\Sigma+1}}\,
\sum_{m=0}^{\infty} 
\frac{
\cQ_{s}( \frac{2\sqrt{\lambda}\, U}{\eta}
+
\frac{2\Sigma - 2}{3};\, \tau ,\, \ell,\, m)
}{
  \frac{2\sqrt{\lambda}\, S}{\eta}
+
\frac{2\Sigma - 2}{3}- \tau  - 2m
} ~.
\qquad
\eeqn
Picking the residue at 
\be
\eta_*=\frac{2\sqrt \lambda S}{\tau+2m-\frac23(\Sigma-1) }~,
\ee
we get
\beqn
A_{\tau,\ell}(S,T)|_{S\text{-pole}} 
 &=&
-\frac{
\Gamma(\Sigma)}{2S}
\int \frac{d\eta}{2\pi i}\,
\frac{e^{\eta}}{\eta^{\Sigma+1}}\,
\sum_{m=0}^{\infty} 
\frac{
\cQ_{s}\!\left( \frac{2\sqrt{\lambda}\, U}{\eta}
+
\frac{2\Sigma - 2}{3};\, \tau ,\, \ell,\, m\right)
}{
  \frac{1}{\eta^*}
-   \frac{1}{\eta}
} 
\\&=&
-\frac{
\Gamma(\Sigma)}{2S} 
\frac{e^{\eta_* }}{\eta_*^{\Sigma-1}}\,
\sum_{m=0}^{\infty} 
\cQ_{s}\!\left( \frac{2\sqrt{\lambda}\, U}{\eta_*}
+
\frac{2\Sigma - 2}{3};\, \tau ,\, \ell,\, m\right)
\\&=&
-\frac{
\Gamma(\Sigma)}{2S} 
\frac{e^{\eta_* }}{\eta_*^{\Sigma-1}}\,
\sum_{m=0}^{\infty} 
\cQ_{s}\!\left( \frac{ U}{ S}(\tau+2m)
+
\frac{  2}{3}(\Sigma-1)\Bigl(1-\frac{ U}{ S}\Bigr);\, \tau ,\, \ell,\, m\right)~.
\qquad\qquad
\eeqn

We are interested in the strong-coupling limit $\lambda\to\infty$ and focus on stringy operators with large twist $\tau\sim \lambda^{1/4}$. 
In this regime, the sum over $m$ can be approximated by an integral over the continuum variable $x=m/\tau^2$ \cite{Alday:2022uxp,Alday:2022xwz}. 
Introducing the rescaled twist $\widetilde{\tau}=\tau/\lambda^{1/4}$ and substituting the explicit expressions for the Mellin blocks, we obtain the following strong-coupling expansion.
\beqn
A_{\tau,\ell}(S,T)|_{S\text{-pole}} &=&
\frac {4^{\ell+\tau }\Gamma(\Sigma) \sin  \frac{   \pi  (\alpha -\tau )}{2}  \sin  \frac{   \pi  (\beta -\tau )}{2}     i^{-\alpha -\beta +2 \Sigma } }
 {\lambda ^{\frac{\Sigma -1}{2}+\frac{3}{4}} (1+\delta_{\ell,0})}
\sum_{i=0}^\infty \frac{ R_{\tau,\ell}^{ i}(S,U)}{\lambda^{\frac{ i}{ 4}}}~.
\qquad
\eeqn
  Note that $ R_{\tau,\ell}^{ i}(S,U)$ also implicitly depends on $p_1,p_2,p_3,p_4$ or equivalently $\alpha, \beta, \gamma, \Sigma$ through \eqref{abgp}. Explicitly, we have 
     \beqn
 R_{\tau,\ell}^{  0}(S,U)&=&
\frac{16 S^{-\Sigma } }{\pi ^3     \left( \widetilde{\tau} ^3-4 S  \widetilde{\tau} \right)}
T_\ell\left(\frac{2 U}{S}+1\right)
\\
 R_{\tau,\ell}^{  1}(S,U)&=&
 -
\frac{ 8S^{-\Sigma }
\Bigl(
4 S\bigl(  2\ell+3 +\alpha^{2}+\beta^{2}\bigr)
+  \bigl( 2\ell+1 +\alpha^{2}+\beta^{2}\bigr)\widetilde{\tau}^{2}
\Bigr)}
{\pi^{3} \bigl( \widetilde{\tau}^{3} -4S\widetilde{\tau}  \bigr)^{2}}\,
T_{\ell}\!\left(1 + \frac{2U}{S}\right)~. \qquad\quad
\eeqn
where $T_\ell(x)$ denotes the Chebyshev polynomials of the $\ell$-th order.     $ R_{\tau,\ell}^{ i}(S,U)$  becomes more complicated as we increase the order $i$.
 Nevertheless, we succeed in finding the closed form expressions  of $ R_{\tau,\ell}^{ i}(S,U)$   for $i\le 3$.

As a result, we have (note that $S+T+U=0$)
\be\label{Acftblock}
\cA(S,T,m_s,m_t)
=\sum_{\tau, \ell} f(\tau, \ell,m_s,m_t) \sum_{i=0}^\infty  \frac{  R_{\tau,\ell}^{ i}(S,U)}{\lambda^{\frac{ i}{ 4}}}+\cdots~,
 \ee
where  $f$ is related to  $\cC_s$   by an overall rescaling
 \be\label{fexpres}
f(\tau, \ell ,m_s,m_t)=
\frac {4^{\ell+\tau }\Gamma(\Sigma) \sin  \frac{   \pi  (\alpha -\tau )}{2}  \sin  \frac{   \pi  (\beta -\tau )}{2}       i^{-\alpha -\beta +2 \Sigma } }
{\lambda ^{\frac{\Sigma -1}{2}+\frac{3}{4}} (1+\delta_{\ell,0})}
 \cC_s(\tau, \ell ;m_s,m_t) ~.
\ee
 
 We can then perform the  following strong coupling expansion for the CFT data of long multiplets:  
\beqn\label{taufexp}
\tau &=& \tau_0  \lambda^{1/4} + \tau_1 + \tau_2  \lambda^{-1/4}  + \tau_3  \lambda^{-3/4} + \cdots ,
\\  \label{taufexp2}
f &=& f_0 + f_1 \lambda^{-1/4} + f_2 \lambda^{-1/2} + f_3 \lambda^{-3/4} + \cdots ,
\eeqn
where  all $\tau_i,f_i$ are independent of the coupling $\lambda$.

Substituting \eqref{taufexp}\eqref{taufexp2} into \eqref{Acftblock}, we get the strong coupling expansion of $\cA$ in the   limit of $\lambda\to \infty$:
\be\label{caord}
\cA(S,T,m_s,m_t)
 \sim \sum_{i=1}^\infty  \frac{\cA_{\text{CFT}}^{(i/2)}|_{S\text{-pole}}}{\lambda^{\frac{ i}{ 4}}}+\cdots~,
\ee
where the notation $\cA_{\text{CFT}}^{(i/2)}|_{S\text{-pole}}$  indicates the contribution  to the $S$-channel poles. This should be compared with   \eqref{Aexpansion}.

We now analyze \eqref{caord} order by order in $\lambda$. 

\bigskip\noindent{\bf $\bullet \;  \cO(\lambda^0)$.} In particular, the leading order $\cO(\lambda^0)$   $S$-pole term in \eqref{Acftblock} is given by 
\be\label{ACFT0}
\cA_{\text{CFT}}^{(0)}|_{S\text{-pole}}\equiv\sum_{\tau_0,\ell}
 f _0  R_{\tau_0 \lambda^{\frac{1}{4}},\ell}^{  0}(S,U) 
 =\sum_{\tau_0,\ell}
 f _0  
 \frac{16 S^{-\Sigma } }{\pi ^3     \left(  {\tau_0} ^2-4 S  \right)  {\tau_0} }
T_\ell\left(\frac{2 U}{S}+1\right)~,
\ee
which has simple pole at $S=\tau_0^2/4$.
This should match the behavior of the  flat space VS amplitude $ \cA^{(0)} $ in \eqref{A0STVS}, which has simple poles at $S = \delta \in \mathbb{Z}_{\geq 0}$.  As a result, we have $\tau_0=2\sqrt{\delta}$ for $\delta\in \bZ_{>0}$. 
\footnote{Note that we are considering the long multiplet whose scaling dimension is large in the strong coupling limit, so $\tau_0$ can not be zero. To match the pole structure at $S=0$, one should consider the short multiplets. }

We can also match the residues at these poles. In particular, from \eqref{A0STVS}, we get
\beqn\label{resA0st}
\Res_{S=\delta} \cA^{(0)}(S,T)
&=&
  -\frac{ \delta ^2+U^2+\delta  U   }{2  }  \left(\frac{(U+1)_{\delta -1}}{\delta !}\right) ^2~,
\eeqn
which is a polynomial of $U$ with degree $2\delta$.  
Similarly from  \eqref{ACFT0}
\beqn\label{Asdel}
\Res_{S=\delta} \cA_{\text{CFT}}^{(0)} |_{S\text{-pole}}
 = \sum_{\ell=0}^{\infty}  
  \frac{-2  f_0(\delta, \ell)}{    \pi ^3 \delta^{\Sigma+\frac12 }  }
  T_\ell\left(\frac{2 U}{\delta}+1\right) ~.
    \eeqn

Comparing the two, we get the    equation
\be\label{ACFTAWS0}
\sum_{\ell=0}^{2\delta}  
  \frac{-2  f_0(\delta, \ell)}{    \pi ^3 \delta^{\Sigma+\frac12 }  }
  T_\ell\left(\frac{2 U}{\delta}+1\right)
  =
    -\frac{ \delta ^2+U^2+\delta  U   }{2  }  \left(\frac{(U+1)_{\delta -1}}{\delta !}\right) ^2, \qquad 
    \delta \in \bZ_{>0}~,
\ee
which should hold for all $U$.  In particular, the upper bound of the  spin $\ell$ is given by $2\delta$ in order to get a polynomial of $U$ with degree $2\delta$. Solving this equation, we can extract the OPE coefficients $f_0$. For example, on the leading  Regge trajectory $\ell=2\delta  $, we find

\be\label{f0evenspin}
 f_0(\delta,2\delta) \equiv 
\frac{\pi^{3}\,\delta^{2\delta + \Sigma - \tfrac{3}{2}}}{2^{4\delta + 1}\,\Gamma(\delta)^{2}} \, .
\ee

Next we have $ f_0(\delta,2\delta-1)=0$. Actually one can easily show that for odd spin 
\footnote{Note that  \eqref{resA0st} can also be written as  $\Res_{S=\delta} \cA^{(0)}(S,T)=-\frac{ \delta ^2+T^2+\delta  T   }{2  }  \left(\frac{(T+1)_{\delta -1}}{\delta !}\right) ^2$, where $T=-U-\delta $. The invariance under the replacement of $U$ by $T$ is just a consequence of crossing symmetry.  Similarly,  one can also  consider the LHS of   \eqref{ACFTAWS0}   and replace $U$ by  $T$. Using the property that $T_\ell(2T/\delta+1)=T_\ell(-2U/\delta-1,\ell)=(-1)^\ell T_\ell(2 U/\delta+1)$, one finds that the odd spin contribution in the LHS of \eqref{ACFTAWS0} gets a negative sign, which implies that the corresponding term must vanish  in order to preserve the invariance of the RHS under the exchange $U\leftrightarrow T$.} 
\be\label{f0oddspin}
f_0(\delta, \ell)=0~, \qquad
\ell \in 2\bN+1~.
\ee

The next two non-trivial results are
 \beqn\label{f0subsub}
  f_0(\delta,2\delta-2)  (1+\delta_{2\delta-2,0})&=& 
\frac{\pi ^3   (    \delta^2 +24\delta  -4) \delta ^{2 \delta +\Sigma -\frac{5}{2}}}{3 \times2^{4\delta } \Gamma (\delta )^2}~,
\\
  f_0(\delta,2\delta-4)  (1+\delta_{2\delta-4,0})&=&
\frac{\pi ^3   (  10 \delta ^5+483 \delta ^4+1360 \delta ^3-2160 \delta ^2+160 \delta +192) \delta ^{2 \delta +\Sigma -\frac{9}{2}}}{45 \times2^{4\delta +1} \Gamma (\delta )^2}~,
\qquad\qquad 
\eeqn
where the factor $(1+\delta_{\ell,0})$  accounts for the special case of $\ell=0$.

\bigskip\noindent{\bf $\bullet \;  \cO(\lambda^{-1/4})$.}   Moving to the next order $\cO(\lambda^{-1/4})$, we have $\cA^{(\frac12) }=0$ \eqref{Afracpw}. This imposes non-trivial constraints on the block expansion in \eqref{Acftblock}
 \beqn
0&=&\cA_{\text{CFT}}^{(\frac12)}|_{S\text{-pole}}\equiv-  \frac{S^{-\Sigma }}{2 \pi ^3 \delta  (S-\delta )^2}\Big(
\delta\, f_0\!\left(\alpha^2+\beta^2+2\ell+6\tau_1+1\right)
+ f_0\, S\!\left(-\alpha^2-\beta^2+2\ell-2\tau_1+3\right)
\nonumber\\
&&\qquad\qquad\qquad\quad  
+ 4\sqrt{\delta}\, f_1\,(S-\delta)
\Big) T_\ell(1+2U/\delta) ~.
 \eeqn
Solving this equation yields \footnote{Rigorously speaking,   if $f_0=0$, $\tau_1$ is not determined but we still have $f_1=0$.}
\be\label{tau1f1}
\tau_1(\delta, \ell)=-1-\ell~, 
\qquad
 \frac{f_1(\delta, \ell) }{  f_0(\delta, \ell)}
=-\frac{5+4\ell-\alpha^2-\beta^2}{4\sqrt\delta}~.
 \ee
 
\bigskip\noindent{\bf $\bullet \;  \cO(\lambda^{-1/2})$.}     Moving further to the next order at $\cO(\lambda^{-\frac12})$ by expanding \eqref{Acftblock}, we find that 
 \be\label{A1cft}
 \cA_{\text{CFT}}^{(1)}|_{S\text{-pole}}=\sum_{\delta }\sum_{i=1}^{4}\frac{ \cW^{\text{CFT}}_i(\delta, U)}{(S-\delta)^i}~, \qquad\qquad
 \cW^{\text{CFT}}_i(\delta, U)=\sum_{\ell=0}^{2\delta} \cW_i(\ell,\delta,  U) ~,
  \ee
where
\footnote{Note that the spectrums may be degenerated, namely several operators may share exactly the same quantum numbers  $\delta,\ell, j, \bar j$ but  have different anomalous dimensions and/or OPE coefficients. Our method can not distinguish the degeneracies, so  our formula $(\tau_i)^s f_j$ should be understood as $\sum_{O} (\tau_i^O)^s  f_j^O$ by summing over all $O$ with the same $\delta, \ell, j,\bar j$ for $s=0,1$.  For simplicity of notation, we will ignore   the summation over $O$. Furthermore, note that the leading Regge trajectory $\ell=2\delta$ is non-degenerated.  }
\beqn\label{R4exp}
 \cW_4(\ell,\delta,U) &=&
-\frac{2 f_0\,\delta^{\frac{3}{2}-\Sigma}\,\mathcal{T}^0_\ell(\delta,U)}{\pi^{3}},
\\[4pt]
 \cW_3(\ell,\delta,U) &=&
-\frac{f_0\,\delta^{\frac{1}{2}-\Sigma}\left[-2(-4+\Sigma)\,\mathcal{T}^0_\ell(\delta,U)
+ 3\delta\,\mathcal{T}_\ell^{1}(\delta,U)\right]}{3\pi^{3}},
\\[6pt]
 \cW_2(\ell,\delta,U) &=&
\frac{f_0\,\delta^{-\frac12-\Sigma}}{6\pi^{3}}
\Big[
\big(-5 - 3\ell^{2} + 3\gamma^{2} + \Sigma(4+\Sigma) - 12\sqrt{\delta}\,\tau_2\big)\,
\mathcal{T}^0_\ell(\delta,U)
\nonumber\\&&\qquad\qquad\quad
+ 4\big(\delta(-1+\Sigma) + U(2+\Sigma)\big)\,\mathcal{T}_\ell^{1}(\delta,U)
+ 6U(U+\delta)\,\mathcal{T}_\ell^{2}(\delta,U)
\Big],\qquad
\qquad \\[6pt] 
 \cW_1(\ell,\delta,U) &=&
\frac{\delta^{-\Sigma - \frac{3}{2}}}{48 \pi^{3}}
\Bigg[
\mathcal{T}^0_\ell(\delta,U)\Big(
f_0 \big(
3(\alpha^4 + 2\alpha^2(\beta^2 - 5) + \beta^4)
- 30\beta^2
- 24\gamma^2\Sigma
\nonumber\\&&\qquad\qquad 
\qquad\quad\quad
+ 48\sqrt{\delta}\,(2\Sigma + 1)\,\tau_2
+ 24\ell^2(\Sigma+1)
- 24\ell(\alpha^2 + \beta^2 - 3)
\nonumber\\&&\qquad\qquad\quad
\qquad\quad\quad 
- 8\Sigma^3
+ 8\Sigma
+ 51
\big)
- 96\delta f_2
\Big)
\nonumber\\&&\quad
 -8 f_0 U^2 (3\delta + 4U)\,\mathcal{T}_\ell^{3}(\delta,U)
- 16 f_0 U \big(2\delta\Sigma + \delta + 3(\Sigma+2)U\big)\,\mathcal{T}_\ell^{2}(\delta,U)
\nonumber\\&&  \quad
+2f_0  \Big[
\delta\big(-3(\alpha+\beta)^2 - 4(\Sigma-1)\Sigma + 3\big)
\nonumber\\&&\qquad\qquad 
+ 6U\big(-\alpha^2 - \beta^2 - 2\gamma^2 + 8\sqrt{\delta}\,\tau_2
+ 2\ell^2 - 2\Sigma(\Sigma+2) - 1\big)
\Big] \mathcal{T}_\ell^{1}(\delta,U)
\Bigg]~.\label{R1exp}
\nonumber \\
\eeqn
 In the expressions above, we have introduced the following notation
\be
\cT_\ell^n(S,U)=\frac{\p^n}{\p U^n} T_\ell\left(1 + \frac{2U}{S}\right)~,
\ee
which implies 
\be
T_\ell\left(1 + \frac{2U}{S}\right)=\cT^0_\ell(S,U)  ~, \qquad
U_{\ell-1}\left(1 + \frac{2U}{S}\right)=  \frac{S}{2\ell}\cT^1_\ell(S,U) ~.
\ee

If we compare the expression \eqref{R4exp} and \eqref{Asdel}, we easily see that
 
 \be\label{Q41}
 \cW^{\text{CFT}}_4( \delta,  U) =
\sum_{\ell=0}^{2\delta} \cW_4(\ell,\delta,  U) = \delta^2 \Res_{S=\delta} \cA^{(0)}(S,U)~.
\ee
Another obvious relation is:
\be\label{Q43}
\cW_3(\ell,\delta,  U) =(-\frac{\Sigma -4}{3 \delta }+\frac12 \p_U) \cW_4(\ell,\delta,  U) ~, \quad
\cW^{\text{CFT}}_3( \delta,  U) =(-\frac{\Sigma -4}{3 \delta }+\frac12 \p_U)  \cW^{\text{CFT}}_4( \delta,  U) ~.
\ee

The expressions $\cW_i$ above depend on CFT data $\tau_i,f_i$, which will be fixed later by comparing with worldsheet ansatz. This will be discussed in the next subsection.

\bigskip\noindent{\bf $\bullet \;  \cO(\lambda^{-3/4})$.}   Finally, we consider the order $\cO(\lambda^{-3/4})$. Again, according to \eqref{Afracpw}, we have $\cA^{(\frac32)}=0$, which also implies that $\cA_{\text{CFT}}^{(\frac32)}|_{S\text{-pole}}=0$. The explicit expression of $\cA_{\text{CFT}}^{(\frac32)}|_{S\text{-pole}}$  can be similarly obtained by expanding \eqref{Acftblock} to higher order, although the computation is more tedious. Nevertheless, we find that the non-trivial constraints are just
\be
  \tau_3 f_0=0  ~,
\ee
and
\beqn\label{f3cof}
f_3&=&
 \frac{ f_2 \left(\alpha ^2+\beta ^2-4 \ell-5\right)}{4 \sqrt{\delta }}
\nonumber\\&&  +
\frac{ f_0 }{192 \delta ^{3/2}}
 \Big[-\alpha ^6+\alpha ^4 \left(17-3 \beta ^2\right)-\alpha ^2 \left(3 \beta ^4-30 \beta ^2+24 \sqrt{\delta } \tau _2+79\right)-\beta ^6+17 \beta ^4
\nonumber \\&& \qquad\qquad \quad 
 -24 \left(\beta ^2-5\right) \sqrt{\delta } \tau _2-79 \beta ^2+48 \ell^3-36 \ell^2 \left(\alpha ^2+\beta ^2-5\right)
\nonumber  \\&& \qquad\qquad\quad
  +12 \ell \left(\alpha ^4+2 \alpha ^2 \left(\beta ^2-5\right)+\beta ^4-10 \beta ^2+8 \sqrt{\delta } \tau _2 +21\right)+111 \Big]~.
\eeqn
In particular, this means for case with $f_0\neq 0$, we have $\tau_3=0$ and $f_3$ is fully determined by $\tau_2,f_2,f_0$ as above.
 
 We will stop at this order, as the next order $\cO(\lambda^{-1})$  involves more CFT data which can not be fixed alone. It may be determiend by   comparing  with the worldsheet ansatz. However, the previous studies showed that all unknown coefficients can be fixed only if one resorts to  some extra input from localization or integrability \cite{Alday:2022xwz}. Since these extra inputs are not available  yet for \AdS$_3$/CFT$_2$, we don't expect to fix all the coefficients at the order $\cO(\lambda^{-1})$.

\subsection{Curvature correction from string worldsheet}
 In this subsection, we will fix the CFT coefficients in the previous subsection, by employing the worldsheet techniques. 
\subsubsection{Worldsheet ansatz}

Our next goal is to find a worldsheet representation of the AdS Virasoro--Shapiro amplitude \(\cA\) for four arbitrary KK modes \(p_i\). As discussed above, in the \(\AdS \times S\) framework, both \(\cM\) and \(\cA\) are functions of the \(m_{ij}\), or equivalently, of the two independent variables \(m_{s}, m_{t}\).   

The  \(\AdS \times S\)  Virasoro--Shapiro amplitude is supposed to satisfy the crossing symmetry, namely it is invariant under the arbitrary simultaneous permutation of $S,T,U$ and $m_s,m_t,m_u$. In particular, 
\be\label{cAcrossing}
\cA^{(1)}(S,T;m_s,m_t) = \cA^{(1)}(S,U;m_s,m_u) = \cA^{(1)}(U,T;m_u,m_t)~.
\ee
Note that the dependence on $\Sigma$ is implicit, which is crossing invariant. 
  
Based on this property, we can now  make the following worldsheet ansatz which makes the crossing symmetry manifest
\be\label{cA1p1234}
\cA^{(1)}(S,T;m_s,m_t)=\cB(S,T; m_s,m_t)+\cB(S,U; m_s, m_u)+\cB(U,T; m_u, m_t) ~,
\ee
where
\be
\cB(S,T; m_s,m_t)=\int \dd^2 z\; |z|^{-2S-2} |1-z|^{-2T-2}\cG(S,T,z; m_s,m_t)~.
\ee
 
We can combine the three terms into a single worldsheet integral, namely 
\be\label{cA1}
\cA^{(1)}(S,T; m_s,m_t)=\int \dd^2 z\; |z|^{-2S-2} |1-z|^{-2T-2}\cG_{\text{tot}}(S,T,z; m_s,m_t)~, 
\ee
where
\beqn
\cG_{\text{tot}}(S,T,z; m_s,m_t)
\nonumber &=&
\cG(S,T,z; m_s,m_t)+|z|^2\cG(S,U,1/z; m_s, m_u)
\\&&
+|1-z|^2\cG(U,T,z/(z-1); m_u, m_t) ~.
\eeqn

To proceed, we need to understand the structure of the worldsheet integrand. Although there is currently no direct way to derive $\mathcal G$ from a microscopic worldsheet formulation, many of its properties have been identified and studied. 
Following \cite{Alday:2023jdk,Alday:2023mvu,Fardelli:2023fyq}, we assume that $\mathcal G$ can be written as a linear combination of a special class of functions known as single-valued multiple polylogarithms (SVMPLs), denoted by $\mathcal L_w(z)$. 
Here the label $w$ is a word in the alphabet $\{0,1\}$, and the weight of the SVMPL is given by the length of $w$. 
For our purposes, we will need weight-three SVMPLs to construct the worldsheet ansatz. Their explicit expressions are collected in Appendix~\ref{apd-SVMPL}.  
To make the crossing properties   manifest, we will use the following equivalent basis 
\be \label{SVMPLbasis}
 \mathcal{L}^s =\Big(\mathcal{L}_{000}^s, \mathcal{L}_{001}^s, \mathcal{L}_{010}^s, \zeta(3)\Big)~,  \qquad \mathcal{L}^a =\Big(\mathcal{L}_{000}^a, \mathcal{L}_{001}^a, \mathcal{L}_{010}^a\Big)~ ,
\ee
where
  \be
 \begin{aligned} & \mathcal{L}_w^s(z)=\mathcal{L}_w(z)+\mathcal{L}_w(1-z)+\mathcal{L}_w(\bar{z})+\mathcal{L}_w(1-\bar{z})~, \\ & \mathcal{L}_w^a(z)=\mathcal{L}_w(z)-\mathcal{L}_w(1-z)+\mathcal{L}_w(\bar{z})-\mathcal{L}_w(1-\bar{z}) ~.
 \end{aligned}
 \ee
The functions $\mathcal{L}_w^s$ ($\mathcal{L}_w^a$) are symmetric (antisymmetric)     under the exchange of $z \leftrightarrow 1-z$.

With this single-value assumption, we can write $\cG$ as linear combination of the 7  basis in \eqref{SVMPLbasis}
\be\label{cGcoeff}
\cG(S,T,z; m_s,m_t)=\sum_{j=1}^4  \cR_j^s(S,T;m_s,m_t)\cL_j^s(z)+\sum_{j=1}^3  \cR_j^a(S,T;m_s,m_t)\cL_j^a(z)~.
\ee

We make the further assumption that the functions $\cR_j^{s/a}(S,T; m_s, m_t)$ are given by homogeneous polynomials of degree two in \(S\) and \(T\), with coefficients given by the linear combinations of 
\be
\{\Sigma m_s, \Sigma m_t, \Sigma m_u, \Sigma, m_s, m_t, m_u\} ~. 
\ee
In total, there are  $(4+3)\times 7\times 3=147$ coefficients to be fixed.

To fix the coefficients, we first observe that \eqref{cAcrossing} is not the full set of crossing symmetry. The remaining crossing symmetry is the invariance under the simultaneous exchange of $S\leftrightarrow T $ and $m_s \leftrightarrow m_t$, namely $ \cA^{(1)}(S,T;m_s,m_t) 
=\cA^{(1)}(T,S;m_t,m_s) $. \footnote{In principle, one can also implement this crossing symmetry in the ansatz \eqref{cA1p1234} by including all terms. }   This can be satisfied if we also impose: 
\be
\cB(S,T ; m_s,m_t)=\cB(T,S; m_t,m_s)~,
\ee
as one can see from \eqref{cA1p1234}. After imposing this constraint, we find that there are only  75 coefficients unfixed.

\subsubsection{Integration over the worldsheet}
To fix the rest of coefficients, we need to evaluate the worldsheet integrals and compare them with CFT.
The worldsheet integration    involving SVMPL takes the following general form
\be
  I_w(S,T)=\int \dd^2 z\; |z|^{-2S-2}|1-z|^{-2T-2}\cL_w(z)~.
\ee
As discussed in \cite{Alday:2023jdk},   the integral  can be evaluated and admits the following expansion
 \be
  I_w(S,T)=\mathtt{polar}_w+\sum_{p,q=0}^\infty (-S)^p (-T)^q \sum_{W \in 0^p \shuffle 1^q \shuffle w}(\cL_{0W}(1) -\cL_{1W}(1))~,
  \ee
  where  $\shuffle$ denotes the   shuffle product, and  $\mathtt{polar}_w$ denotes the singular terms in $S,T$.
  
The polar terms arise from specific regions of the complex $z$-plane. In particular, all singular terms of the form $\#/(S - \delta)^i$ with $i = 1,2,3,4$ originate from the region where $|z| \to 0$. To isolate these contributions, we expand the integrand around small $|z|$ using polar coordinates $z = \rho e^{i\theta}$ and perform a Taylor expansion in $\rho$ near $\rho = 0$. Integrating term by term over $\theta$ and then over $\rho$, we obtain contributions of the form $\rho^{-1 - 2S + 2\delta} \log^p \rho^2$. The integral evaluates as \cite{Alday:2023jdk,Alday:2023mvu}
  \be\label{polarcont}
  \int_0^ {\rho_0}  \dd \rho\;   \rho^{-2S+2\delta -1} \log^p \rho^2  =-\frac{p!}{2}\frac{1}{(S-\delta)^{p+1}}+ \cdots~,
   \ee
where the dependence on the cutoff $\rho_0$ only enters the regular terms.     Since we consider weight-3 basis in \eqref{SVMPLbasis} as the integrand, the maximal value of $p$ is 3, which implies that maximal order the pole is 4 from \eqref{polarcont}. This is consistent with the expansion in   \eqref{A1cft}, where the  maximal order  of the  poles    is also 4.

  \subsubsection{Fixing the coefficients}
  More  precisely, the equation \eqref{polarcont} allows us to  evaluate the worldsheet integrals   and compute the singular terms    explicitly. For the $S$-channel, one finds 
      \be 
 \cA^{(1)}\sim \sum_{\delta }\sum_{i=1}^{4}\frac{\cW^{\text{WS}}_i(\delta, U)}{(S-\delta)^i}~,  
   \ee
where the explicit expression  of  $\cW^{\text{WS}}_i(\delta, U)$ can be obtained for different values of $\delta\in \bZ_{>0}$, and it depends on the unknown coefficients and $U, m_s,m_t, p_i$.

The consistency of CFT block expansion and worldsheet representation requires that  $\cW_i^{\text{CFT}}(\delta, U)=\cW_i^{\text{WS}}(\delta, U)$ for  $i=1,2,3,4$ and $\delta \in \bZ_{>0}$.   In particular,  \eqref{Q43} and \eqref{Q41} should also hold for  $\cW_i^{\text{WS}}$.
Finally, we should consider the pole of $\cA^{(1)}$ at $S=T=0$, which should be consistent the supergravity result in \eqref{A1SG}.

By evaluating the   polar contributions to the worldsheet integral \eqref{cA1}, and comparing them with  corresponding  CFT expression for  $S$-channel poles at $S = \delta = 1,2,\ldots$ as well as  supergravity   results  \eqref{A1SG} for pole  $S=T=0$, 
it turns out that we are able to  fully determine all physical  coefficients in the worldsheet ansatz, and thereby fix the CFT data at this order.

The final expression of the worldsheet integrand is fully determined  up to 12 ambiguities  whose integrals vanish. These ambiguities do not affect the physical results. For a certain set of choices of those ambiguities, we find the coefficients in the worldsheet ansatz in \eqref{cGcoeff} are given by

\beqn\renewcommand{\arraystretch}{1.5}
\label{RRscof}
\cR^s=\footnotesize
\begin{pmatrix}
\frac{1}{192} \left(2 S T (2 + 9 (m_s  + m_t )(-1 + \Sigma) + 20 \Sigma - 4 \Sigma^2) + 3 (S^2 +T^2 )(5   - 3 \Sigma + \Sigma^2)  - 6( m_s T^2+m_t S^2)(\Sigma-1)\right) \\
\frac{1}{48} \left(-3 m_t S^2 (-1 + \Sigma) - T (3 m_s T (-1 + \Sigma) + S (7 + \Sigma + \Sigma^2))\right) \\
\frac{1}{92} \left(6 T^2 (-2 + m_s (-1 + \Sigma)) + 6 S^2 (-2 + m_t (-1 + \Sigma)) - S T (19 + \Sigma + \Sigma^2)\right) \\
\frac12 T^2 (2 + m_s - m_s \Sigma) + \frac12 S^2 (2 + m_t - m_t \Sigma)  
\end{pmatrix}~,\qquad\quad 
\eeqn

  \beqn\renewcommand{\arraystretch}{1.5}
  \label{RRacof}
  \cR^a=\footnotesize
\begin{pmatrix}
\frac{1}{64} \left(6 (m_s - m_t) S T (-1 + \Sigma) + T^2 (-5 + 2 m_s (-1 + \Sigma) + 3 \Sigma - \Sigma^2) + S^2 (5 - 2 m_t (-1 + \Sigma) - 3 \Sigma + \Sigma^2)\right) \\
\frac{1}{48} \left(-6 (m_s - m_t) S T (-1 + \Sigma) - T^2 (2 + 3 m_s (-1 + \Sigma) + 4 \Sigma) + S^2 (2 + 3 m_t (-1 + \Sigma) + 4 \Sigma)\right) \\
\frac{1}{48} \left(-3 (m_s - m_t) S T (-1 + \Sigma) + S^2 (1 + 2 \Sigma) - T^2 (1 + 2 \Sigma)\right)
\end{pmatrix}~.\qquad\quad 
\eeqn

With the explicit  solution, we can compute the low energy expansion of the $\AdS\times S$ Virasoro-Shapiro amplitude:
  \beqn
  \cA^{(1)}(S,T;m_s,m_t)&=&
  \frac{1-\Sigma}{12}   \left(\frac{-3 m_s+\Sigma -4}{S^2}+\frac{-3 m_t+\Sigma -4}{T^2}+\frac{-3 m_u+\Sigma -4}{U^2}\right)
   \nonumber \\&&
  +\frac{1}{2} (\Sigma -1)  (m_s S+m_t T+m_u U)\zeta (3) 
  \nonumber
  \\&&
 + \Big[  -\frac{1}{12} \left(4 \Sigma ^2+41\right) 
    \left(S^3+T^3+U^3\right) 
    \nonumber    \\&&\qquad
      +(\Sigma -1)   \left(m_s S^3+m_t T^3+m_u U^3\right)\Big]\zeta (5)+\cdots ~.\qquad
  \eeqn
  Actually   the same result is also obtained without choosing any ambiguities, whose roles disappear in the final physical $\AdS\times S$ Virasoro-Shapiro amplitude as expected. The high energy limit is also studied in \cite{Jiang:2025oar}.

  The worldsheet representation and $\AdS\times S$ Virasoro-Shapiro  amplitude above reproduce exactly the results in \cite{Jiang:2025oar}. There the results were  obtained by first studying the special case of $\EV{pp11}$, and then use $\AdS\times S$ formalism to extend to general $\EV{p_1p_2p_3p_4}$. That method gives a shortcut to derive the $\AdS\times S$ Virasoro-Shapiro amplitude, but it is not obvious that the worldsheet representation is fully consistent with the underlying superconformal block expansion. Our derivations in this paper consider the   block expansion  for   external operators with general scaling dimension, and the   remarkable consistency  in the comparison between CFT block expansion and worldsheet  ansatz provides a strong support. 
  
  \subsubsection{CFT data in internal Mellin space}
  The current method also generates lots of  CFT data. For example, on the leading trajectory with $\ell=2\delta$, we find~\footnote{Specializing to the case of $\EV{pp11}$ and $\EV{p11p}$, it is easy to verify that we reproduce the $S$ and $T$-channel results in \cite{Jiang:2025oar}.
} 
  \beqn\label{tau2eq}
  \tau_2(\delta, 2\delta)&=&\frac{\gamma ^2+6 \delta ^2-2 \delta -2  {m_s} (\Sigma -1)+\Sigma ^2-2 \Sigma +1}{4 \sqrt{\delta }}~,
  \\ \label{f2eq}
   f_2(\delta, 2\delta)&=&
   \frac{\pi ^3 4^{-2 \delta -3} \delta ^{2 \delta +\Sigma -\frac{5}{2}}}{3 \Gamma (\delta )^2}
\Big[63 +8 \Sigma  +
3 \left(\alpha ^4+2 \alpha ^2 \left(\beta ^2-5\right)+\beta ^4\right)
-30 \beta ^2+12 \gamma ^2+208 \delta 
\nonumber\\& &\qquad\qquad\qquad\qquad
-4 \left(
18 \alpha ^2 \delta +18 \beta ^2 \delta +28 \delta ^3-12 \delta ^2 \Sigma -78 \delta ^2
+36 \delta  \Sigma  
-4 \Sigma ^3+9 \Sigma ^2
\right)
\nonumber\\& & \qquad\qquad\qquad\qquad
-24 m_s (\Sigma -1) (4 \delta +2 \Sigma +1)
+192 \delta ^3 \zeta (3)
\Big]
\\&=&
   \frac{f_0(\delta,2\delta)}{96 \delta}
\Big[63 +8 \Sigma  +
3 \left(\alpha ^4+2 \alpha ^2 \left(\beta ^2-5\right)+\beta ^4\right)
-30 \beta ^2+12 \gamma ^2+208 \delta
\nonumber\\& &\qquad\qquad\qquad\qquad
-4 \left(
18 \alpha ^2 \delta +18 \beta ^2 \delta +28 \delta ^3-12 \delta ^2 \Sigma -78 \delta ^2
+36 \delta  \Sigma  
-4 \Sigma ^3+9 \Sigma ^2
\right)
\nonumber\\& & \qquad\qquad\qquad\qquad
-24 m_s (\Sigma -1) (4 \delta +2 \Sigma +1)
+192 \delta ^3 \zeta (3)
\Big]~.
  \eeqn
  
 Using \eqref{f3cof}, we can  even compute the CFT data to the next order, yielding   $\tau_3(\delta, 2\delta)=0$ and 
\begin{align}
  \frac{f_3(\delta, 2\delta)}{ f_0(\delta, 2\delta)}
 =& 
 \frac{1}{384 \delta ^{3/2}}
 \Bigg[-33+\alpha^{6}+\beta^{6}
+\alpha^{4}\!\left(-11+3\beta^{2}-48\delta\right)
-560\delta-480\,m_s\,\delta
 \nonumber\\[2pt] &\qquad \qquad
-1616\delta^{2}-768\,m_s\,\delta^{2}
-592\delta^{3}+896\delta^{4}
-\beta^{4}\!\left(11+48\delta\right) \nonumber\\[2pt]
&\qquad\qquad
+\beta^{2}\!\left(43+352\delta+96\,m_s\,\delta+528\delta^{2}-112\delta^{3}\right)
 \nonumber\\[2pt] &\qquad \qquad
+\alpha^{2}\!\left(43+3\beta^{4}+352\delta+96\,m_s\,\delta+528\delta^{2}-112\delta^{3}-6\beta^{2}(5+16\delta)\right) \nonumber\\[2pt]
&\qquad\qquad
+16\left(-5+\alpha^{2}+\beta^{2}-8\delta\right)\!\left(2+m_s(3-6\delta)-9\delta+3\delta^{2}\right)\Sigma
 \nonumber\\[2pt] &\qquad \qquad
-48(1+m_s)\left(-5+\alpha^{2}+\beta^{2}-8\delta\right)\Sigma^{2}
+16\left(-5+\alpha^{2}+\beta^{2}-8\delta\right)\Sigma^{3}
 \nonumber\\[2pt] &\qquad \qquad
+192\left(-5+\alpha^{2}+\beta^{2}-8\delta\right)\delta^{3}\zeta(3)
\Bigg].
\end{align}

  Let us also provide the OPE coefficient on the Regge trajectory with $\ell=2\delta-1$:
  \be\label{f2subRg}
   f_2(\delta, 2\delta-1)=   f_0(\delta, 2\delta)  \Big( -\alpha  \beta -2  {m_s} (\Sigma -1)-4  {m_t} (\Sigma -1)+2 \Sigma ^2-6 \Sigma +4\Big)~.
  \ee
  Note that  $\tau_2   (\delta, 2\delta-1)$ can not be determined because in \eqref{R1exp},  $\tau_2 f_0$ appears as a combination, but $f_0  (\delta, 2\delta-1)=0$ \eqref{f0oddspin}.  Note that $f_1(\delta, 2\delta-1)=0$ also vanishes due to  \eqref{tau1f1}. So $f_2$ is the first non-vanishing order, while  the next order can also be derived from \eqref{f3cof} 
  \be
  f_3(\delta, 2\delta-1)= 
 \frac{    \alpha ^2+\beta ^2-4 \ell-5 }{4 \sqrt{\delta }}
  f_2(\delta, 2\delta-1)~.
  \ee

    Finally let us also give the anomalous dimension   on the Regge trajectory with $\ell=2\delta-2$:~\footnote{The factor  $\delta   ^2 +24\delta-4 $  in the denominator  looks a bit unusual, but this is exactly the factor in \eqref{f0subsub}. Note that $f_0$ and $\tau_2$ appear together in our block expansion. }
  \be\label{tau2subsubRg}
  \tau_2(\delta, 2\delta-2)=  \frac{\gamma ^2+6 \delta ^2-\frac{2 \delta }{3}+\frac{3020 \delta -504}{\delta   ^2 +24\delta-4}-2  {m_s} (\Sigma -1)+\Sigma ^2-2 \Sigma -123}{4 \sqrt{\delta }}~.
  \ee

\section{D1-D5 CFT data and internal spaces} \label{MelSpin}

In the previous sections, we have obtained abundant CFT data from bootstrapping $\AdS_3\times S^3$ Virasoro-Shapiro amplitude, including   anomalous dimensions and OPE coefficients. 
However, these CFT data are derived as functions of the internal Mellin variables, making the physical meaning obscure. The goal of this section is to translate those CFT data from internal Mellin space to internal  spin space, and obtain the physical CFT data as functions of the spacetime conformal weights and   spins under  internal R-symmetry.  We will provide  the explicit procedures of translating between    internal Mellin space to internal  spin space, and also illustrate in some examples. 

\subsection{Spins of exchanged long multiplets }
 
We would like to consider the super-conformal block expansion in the $S$-channel. We are mainly interested in the exchange of long-multiplets,   whose  sueprconformal primaries have quantum numbers $(h,\bar h, j , \bar j)$ under the symmetry $SL(2,\bR)_L\times SL(2,\bR)_R \times SU(2)_L \times SU(2)_R$. These quantum number are also equivalently represented as $\Delta, \ell, a,b$ via
 \be
 h+\bar h =\Delta =\tau+\ell~,  \qquad h-\bar h =\ell~, \qquad \tau=2\bar h~, \qquad \tau+2\ell =2h~,
 \ee
 \be
2\bar j=b~, \qquad 2  j=b+2a~ , \qquad a=  j- \bar j~.
 \ee
 The spacetime and internal spins are quantized: $\ell\in \bZ, j\in \frac12\bZ, \bar j \in \frac12 \bZ$ 
 and $j-\bar j\in \bZ$, as we will see below in \eqref{jjbarsL}. In the parity even case where the system is left-right symmetric, we can then restrict to the quantum numbers $\ell, a\ge 0$.

 Once we fix the external operators, then the exchanged operator in the $S$-channel should have R-symmetry spin $(j,\bar j)$ subject to the condition: 
 \be\label{jjbarsL}
j_{\text{min}}\le  j, \bar j \le j_{\text{min}}+  L-1 ~,\qquad j-j_{\text{min}}\in \bZ ~,\quad \bar j -j_{\text{min}}\in \bZ~, \quad j_{\text{min}}=\frac12\max(|p_1-p_2|,|p_3 -p_4|)~,  
\ee
where     the extremality
$
L=\min_i(p_i,\Sigma -p_i)
$. The lower bound in \eqref{jjbarsL} just follows from the standard Clebsch–Gordan decomposition rule for $SU(2)$, while the upper bound will be tested and justified later. 
 Note that for a physically legitimate four-point correlator, we should have $L\ge 1$. Also note $\Sigma$ must be an integer.    
 
 \subsection{Superconformal block expansion}\label{superbkexp}
 Given an exchange long-multiplet, its contribution to four-point correlation function  is given by the superconformal   block.
 
Let  us first recall   the standard two-dimensional conformal block 
 \footnote{
In the case that $h(\ell)=h(-\ell)$, one has $\sum_{\ell\in \bZ} h(\ell) =2\sum_{\ell\in \bN}h(\ell)/(1+\delta_{\ell,0})$.  The restriction of   summing $\ell$ only over $\bN$ accounts for the factor $1+\delta_{\ell,0}$   in \eqref{Grstl}.  The same comment applies to $a$.
}
 \be\label{Grstl}
G_{\tau, \ell}^{(r,s)}(z,\bar z)
=\frac  {\kappa^{(r,s)}_{\tau  }(z)\kappa^{(r,s)}_{\tau+2\ell}(\bar z)
+
\kappa^{(r,s)}_{\tau +2\ell }(z)\kappa^{(r,s)}_{\tau }(\bar z)
 }
{1+\delta_{\ell,0}}
=\frac  {\kappa^{(r,s)}_{2h  }(z)\kappa^{(r,s)}_{ 2\bar h}(\bar z)
+
\kappa^{(r,s)}_{2\bar h}(z)\kappa^{(r,s)}_{2h }(\bar z)
 }
{1+\delta_{h,\bar h }}~,
 \ee
 where  
  \be
   \kappa^{(r,s)}_\beta(x) =x^{\beta/2}{}_2F_1( \frac{\beta+r}{2},\frac{\beta+s}{2};  \beta;x)~.
 \ee 
Due to the  invariance under the exchange of $z$ and $\bar z$, $G$ is actually a function of $U,V$. 
From the  second expression in \eqref{Grstl}, it is also obvious to see that     $G$  is symmetric under the exchange of $h, \bar h$.

Looking at the four-point function  \eqref{cOKG} again,  we can decompose    $\wt\cG$  as follows
\be\label{cgtilde}
\wt \cG =\text{protected}+(z-y) (\bar z -\bar y)  \wt \cH(U,V; \sigma, \rho)
=\text{protected}+y \bar y(1-z/y ) (1-\bar z /\bar y) \wt \cH(U,V; \sigma, \rho)~,
\ee
 which gives the definition of the reduced correlator $\wt \cH$. 
Then  we can expand the  reduced correlator $\wt\cH$ in terms of the    superconformal block for  exchagned long-multiplets as follows~\cite{Aprile:2021mvq}
 \be
 \wt\cH =\sum_{\tau, \ell,a,b} \sfC_s({\tau, \ell,a,b} )\wt\cH_{\tau, \ell,a,b} ~,
 \ee
 where \footnote{Note that we have removed the overall factor $\frac14$ in \cite{Aprile:2021mvq} for simplicity. }
  \be
\wt\cH_{\tau, \ell,a,b} = 
(-1)^a \frac{ G_{-b, -a}^{(p_{12}, p_{43})}(y,\bar y) }{y \bar y }
  \frac{G_{\tau+2, \ell}^{(-p_{12},-p_{43})}(z ,\bar z) }{z\bar z} ~.
   \ee
 Note that in the above expression $\tau$ is also a function of $\delta,\ell,a,b$ and $\lambda$, and the definition of $G_{-b, -a}^{(p_{12}, p_{43})}(y,\bar y)$ is 
   also given by \eqref{Grstl} up to replacements of various symbols. In particular, $G_{-b, -a}^{(p_{12}, p_{43})}(y,\bar y)$  is invariant under the exchange of $y\leftrightarrow \bar y$, so it is actually a function of $\sigma, \rho$.

 The reduced correlator $\wt\cH$ is related in a simply way to the previous reduced correlator $  \cH$ in  \eqref{cgs}. By comparing \eqref{cgtilde} and \eqref{cgs} and using the relation \eqref{KKtdrel}, we find that $\cH$ can be similarly expanded as

 \be\label{cHs}
\cH =\sum_{\tau, \ell,a,b} \sfC_s({\tau, \ell,a,b} ) \cH_{\tau, \ell,a,b} ~,
 \ee
 where
  \beqn
 \cH_{\tau, \ell,a,b} &=&
 y\bar y \Big(\frac{ V}{U\rho} \Big) ^{\frac{p_3+p_4-2L}{2}}
 \Big(\frac{\rho }{V\sigma} \Big) ^{\frac{p_1-p_2}{2}} \wt\cH_{\tau, \ell,a,b}
 \\ &=&
(-1)^a \Big(\frac{ V}{U\rho} \Big) ^{\frac{p_3+p_4-2L}{2}}
 \Big(\frac{\rho }{V\sigma} \Big) ^{\frac{p_1-p_2}{2}}   { G_{-b, -a}^{(p_{12}, p_{43})}(y,\bar y) }   \frac{G_{\tau+2, \ell}^{(-p_{12},-p_{43})}(z ,\bar z) }{U}  
  \\ &=& 
E_{\tau+2, \ell}^{(-p_{12},-p_{43})}(U,V)
  Z_{-b,-a}^{( p_{12}, p_{43})}(\sigma,\rho) ~,
 \eeqn
and 
\beqn\label{EUV}
E_{\tau+2, \ell}^{(-p_{12},-p_{43})}(U,V)
&= & U^{L-1-\frac{p_3+p_4}{2}}  V^{-L +\frac{-p_1+p_2+p_3+p_4}{2}}
   {G_{\tau+2, \ell}^{(-p_{12},-p_{43})}(z ,\bar z) } ~,
\\ \label{Rblock33}
Z_{-b,-a}^{( p_{12}, p_{43})}(\sigma,\rho)
& =&
 (-1)^a \Big(\frac{ 1}{ \rho} \Big) ^{\frac{p_3+p_4-2L}{2}}
 \Big(\frac{\rho }{ \sigma} \Big) ^{\frac{p_1-p_2}{2}}   G_{-b, -  a}^{(p_{12}, p_{43})}(y,\bar y) ~.
 \eeqn 
 From the expressions above,  the superconformal block is given by the product from spacetime and internal part. We will focus on the internal part, as the spacetime part gives the the Mack polynomial  \eqref{melblock} after performing the Mellin transformation \eqref{MellinExt}, which has been discussed before.

 One can verify that the block \eqref{Rblock33} has all the expected properties. To illustrate this, let us assume that  $p_1 \ge  \{p_2 , p_3\} \ge p_4 $. Then $L$ is given by 
\be\label{Lp1234}
 L 
=\frac{p_4+p_3}{2}-\frac12 \max(p_ {34},p_ {12})=\begin{dcases} 
p_4 ~,& \text{if}\ p_4-p_3\le p_2-p_1~,\\
\frac{p_2+p_3+p_4-p_1}{2} ~, &\text{if}\ p_4-p_3\ge p_2-p_1~. \\
\end{dcases}
 \ee
One can then verify that   \eqref{Rblock33}
  is indeed a function of $\sigma, \rho$ with degree $\max(j,\bar j)-j_{\text{min}}$ for $j,\bar j \ge j_{\text{min}}$ and $j-j_{\text{min}},\bar j-j_{\text{min}}\in \bN$, where $j_{\text{min}}=\frac12\max(|p_1-p_2|,|p_3 -p_4|)$. Further imposing  the selection rule \eqref{jjbarsL}, one finds that  \eqref{Rblock33} is
  indeed a polynomial in $\sigma, \rho$ of degree at most $L-1$. 
  In the most special case $j=\bar j=j_{\text{min}}$, one can verify that $Z= 1 $.~\footnote{Let us denote $r=p_{12},s=p_{43}$. Note that $r>0,s<0$.  There are two possibilities. If $r \le -s$, then we have $L=p_4$, $j_{\text{min}}=-\frac12s$, and
    \be\nonumber
Z_{-2j_{\text{min}},0}^{( p_{12}, p_{43})} =Z_{  s,0}^{( r,s)} =
 \Big(\frac{ 1}{ \tau} \Big) ^{\frac{p_3+p_4-2L}{2}}
 \Big(\frac{\tau }{ \sigma} \Big) ^{\frac{p_1-p_2}{2}}   y^{ \frac s 2} (1-y)^{-\frac{r+s}{2}}\bar y^{ \frac s 2} (1-\bar y)^{-\frac{r+s}{2}}   =1~.
 \ee
If $r\ge -s$, then $L=\frac{p_2+p_3+p_4-p_1}{2}$, $j_{\text{min}}=\frac12 r$ and 
 \be\nonumber
Z_{-2j_{\text{min}},0}^{( p_{12}, p_{43})} =Z_{  -r,0}^{( r,s)} =
 \Big(\frac{ 1}{ \tau} \Big) ^{\frac{p_3+p_4-2L}{2}}
 \Big(\frac{\tau }{ \sigma} \Big) ^{\frac{p_1-p_2}{2}}   y^{- \frac r 2} \bar y^{ -\frac r 2}  =1  ~.
 \ee
In the above computations, we used \eqref{yybrel} and the following identities: 
  \be\label{krsval} \nonumber
   \kappa^{(r,s)}_ {-r }(x) =x^{-\frac r 2} ~, \qquad
      \kappa^{(r,s)}_ {-s }(x) =x^{-\frac s 2}~,  \qquad
        \kappa^{(r,s)}_ { r }(x) =x^{ \frac r 2} (1-x)^{-\frac{r+s}{2}}~, \qquad
      \kappa^{(r,s)}_ { s }(x) =x^{ \frac s 2} (1-x)^{-\frac{r+s}{2}}~.
 \ee 
   }
 This in particular guarantees that the reduced correlator $\cH$, as a linear combination of $Z$  in \eqref{cHs},  is  also a polynomial in $\sigma, \rho$     of degree $L-1$ \cite{Rastelli:2019gtj}, as expected.

Therefore, we find the following superconformal block expansion for $\cH$:
\be
\cH (U,V;\sigma, \rho)=\sum_{\tau, \ell,a,b} \sfC_s({\tau , \ell,a,b} ) E_{\tau+2, \ell}^{(-p_{12},-p_{43})}(U,V)
  Z_{-b,-a}^{( p_{12}, p_{43})}(\sigma,\rho) ~.
 \ee
 
Performing the external Mellin transform   in \eqref{MellinExt}---more precisely  its inverse Mellin transform---we arrive at
\be\label{Mst2}
M(s,t,\sigma,\rho)
=
\sum_{\tau ,\ell}
C_{s}(\tau ,\ell;\sigma,\rho)
\sum_{m=0}^{\infty}
\frac{
\cQ_{s}(\tilde{u};\, \tau ,\, \ell,\, m)
}{
s - \tau  - 2m
}
\;+\;\cdots, \qquad   
\ee
 and the coefficient  $C_s$ is
 \be\label{Csrhosigma}
 C_{s}(\tau ,\ell;\sigma,\rho)=\sum_{a,b}\sfC_s({\tau , \ell,a,b} )
  Z_{-b,-a}^{( p_{12}, p_{43})}(\sigma,\rho) ~,
 \ee
 where the summation over $a,b$ or equivalently $j, \bar j$ is subject to the condition \eqref{jjbarsL}. 
 
 Further performing the  internal Mellin transformation  \eqref{MellinInt} yields the previous expansion \eqref{Mst}, where the coefficient $\cC_s$ is related to $C_s$ above via
  \be\label{Msmij}
 C_{s}(\tau ,\ell;\sigma,\rho)=\sum_{m_{24},m_{34} }
\frac{   \sigma^{m_{24}}\rho^{L-1-m_{24}-m_{34}}  
}{  \Gamma_S(m_s,m_t) }
 \cC_{s}(\tau ,\ell,m_s,m_t)~.
\ee
 
 
One can readily check that there are in total $L(L+1)/2$ terms in the sum on the right-hand side of \eqref{Msmij}.~\footnote{One way to see this is to consider the ordering $p_1 \ge \{p_2,p_3\} \ge p_4$.}
 Accordingly, $\mathcal{C}_{s}(\tau,\ell; m_{ij})$ involves $L(L+1)/2$ independent coefficients. This matches the number of allowed internal spins for the exchanged operators, as can be seen from \eqref{jjbarsL}.  
 \footnote{Note that $(j,\bar j)$ and $(\bar j,j)$ are related by parity, so they are not independent.}
  
 Eq.~\eqref{Csrhosigma} and \eqref{Msmij} allow  us to  establish the   connection between $\sfC_s$ and $\cC_s$.  To fulfill this, one can  apply the   Mellin transformation \eqref{Msmij} to the   block \eqref{Rblock33}. We first note that \eqref{Rblock33} can be simplified: 
  \be\label{Rblock2}
Z_{-b,-a}^{( p_{12}, p_{43})}(\sigma,\rho) =
  {  }{ \rho} ^{-\frac{p_3+p_4 }{2}+L}   G_{-b, -  a}^{(p_{12}, p_{34})}
  \Big(\frac{y}{ y-1 },\frac{\bar y }{ \bar y-1}\Big) ~.
 \ee
 Note that the exchange of $y\leftrightarrow y/(y-1)$ is equivalent to exchange $\sigma\leftrightarrow \rho$,
where we used the Pfaff transformation  of hypergeometric function
 \be
 {}_2F_1(a,b;c;z) = (1-z)^{-a}\,{}_2F_1\!\left(a,\,c-b;\,c;\,\frac{z}{z-1}\right)~,
 \ee
 and consequently 
 \be
 \kappa_\beta^{(r,s)}(x)= e^{\pm \frac12 \pi i \beta}(1-x)^{-\frac12 r}\kappa_\beta^{(r,-s)}(\frac{x}{x-1})~.
 \ee

Physically, it is more convenient to label the operators with internal SU(2) spins $j,\bar j$, so we define
   \beqn\label{Rblock}
\hat Z_{j,\bar j}^{( p_{12}, p_{34})}(  \rho,\sigma) 
&=&
\frac{(1+\delta_{a,0}) }{2} Z_{-b,-a}^{( p_{12}, p_{43})}(    \sigma,\rho)|_{\rho\leftrightarrow\sigma  }
=\frac12
(1+\delta_{a,0}) { \sigma} ^{-\frac{p_3+p_4 }{2}+L}   G_{-b, -  a}^{(p_{12}, p_{34})}(y,\bar y ) 
\nonumber \\&=&
\frac12 \sigma ^{-\frac{p_3+p_4 }{2}+L}  
 \Big(
  \kappa^{( p_{12}, p_{34})}_{-2j  }(y)\kappa^{( p_{12}, p_{34})}_{ -2\bar j}(\bar y)
+
  \kappa^{( p_{12}, p_{34})}_{-2j  }(\bar y)\kappa^{( p_{12}, p_{34})}_{ -2\bar j}(  y)
  \Big)~,
  \label{Rblock2}
  \qquad\qquad
  \eeqn
  where $a,b $ are related to $j,\bar j$ via $  2\bar j=b,\; a=  j- \bar j$.

 Similar to \eqref{Msmij}, we can consider the internal Mellin transform for $\hat\cZ$
 
 \be\label{Zseq232}
 \hat Z_{j,\bar j}^{( p_{12}, p_{34})}(  \rho,\sigma)  
 =\sum_{m_{24},m_{34} }
\frac{   \rho^{m_{24}}\sigma^{L-1-m_{24}-m_{34}}  
}{  \Gamma_{S} (m_s,m_t)}
\hat \cZ ^{( p_{12}, p_{34})} (j,\bar j; m_{s},m_t)~.
\ee
where we exchanged the role of $\rho, \sigma$, in consistency with \eqref{Rblock2}.

Eq.~\eqref{Csrhosigma} then becomes
 \be
    C_{s}(\tau ,\ell;  \rho,\sigma)=\sum_{j\ge \bar j}   \sfC_s({\tau, \ell; j, \bar j } )
\hat Z_{j,\bar j}^{( p_{12}, p_{34})}(  \rho,\sigma)~,\qquad
 \sfC_s({\tau, \ell; j, \bar j } )= 2  \sfC_s({\tau, \ell,a,b } )/(1+\delta_{a,0})~.
 \ee
Comparing them with Eq.~\eqref{Msmij} yields
 
 \be\label{Cstaulm}
  \cC_{s}(\tau ,\ell;m_s,m_t)=\sum_{j\ge \bar j} \sfC_s({\tau, \ell; j, \bar j } )
\hat   \cZ_{j, \bar j }^{( p_{12}, p_{34})} ( m_{s},m_t)~.
 \ee

Our ultimate goal is to compute  $   \sfC_s({\tau, \ell; j, \bar j } )$.
 Since we have bootstrapped  $\cC_s$ on the LHS in previous sections, the only remaining task is to figure out the  function $\hat \cZ $, which is the  internal space analogue of the Mack polynomial.   
 The derivation of $\hat \cZ$  is presented in Appendix~\ref{MackS}, and the final result is given by Eq.~\eqref{zhatjm} there:
   \beqn \label{zhatjms}
\hat \cZ 
( j, \bar j;m_{s},m_{t})&=& 
 \sum_{n,\bar n    }
 \frac{(j- \alpha/2)!\,(j- \beta/2)!}{(n- \alpha/2)!\,(n- \beta/2)!}\,  
  \frac{(\bar j- \alpha/2)!\,(\bar  j- \beta/2)!}{(\bar  n- \alpha/2)!\,(\bar  n- \beta/2)!}\,  
 \frac{ (j+n  )! (-1)^n}{(j-n)!(2j)!}
 \frac{ (\bar j+\bar n  )! (-1)^{\bar n}}{(\bar j-\bar n)!(2\bar j)!} 
 \nonumber \\&&   \qquad 
\times 
\Theta_{-\bar n,-n}
\Big({\frac{2m_t+\alpha+\beta}{4},\frac{2\Sigma -4-2m_s-2m_t-\alpha-\beta  }{4}}\Big)
       \Gamma_{S}(m_{s},m_{t})~,
\label{zhatjmeq}
\nonumber
\\
 \eeqn
 where  we have suppressed the superscript $ {(\alpha ,\beta)} \equiv (  p_1-p_2,p_3-p_4)$  in order to ease the notation. Note that   the summation over  $n =\frac12\max(\alpha,\beta) ,  \cdots, j$ and $\bar n =\frac12\max(\alpha,\beta) ,\cdots,\bar j$ is enforced automatically by the factorial function in the denominator.  The $\Theta$ function is defined via  
  \be\label{ybtheta}
y^m \bar y^n+y^n \bar y^m
=2\sum_{p,q}\Theta_{m,n}({p,q}) \sigma^p \rho^q~,
\ee
and the explicit expression of $\Theta$ is given in Eq.~\eqref{thetamnpq}.

This formula \eqref{zhatjmeq} holds provided $\alpha \equiv p_1-p_2 \ge 0$ and $\beta \equiv p_3-p_4 \ge 0$.
We can always relabel operators $1\leftrightarrow 2$ and/or $3\leftrightarrow 4$ to satisfy these conditions without changing  the $S$-channel CFT data. 
 We can regard $\hat \cZ$ as a  square matrix   of size $L(L+1)/2$ 
  and  define  its inverse as follows 
  \beqn
\sum_{j,\bar j}     \hat \cZ^{-1}  (  m'_{s},m'_{t};j, \bar j )   \cZ  ( j, \bar j;m_{s},m_{t})  
&=& 
\delta_{m_s,m_s'}\delta_{m_t,m_t'}~,
\\
\sum_{m_s,m_t}     \hat \cZ  ( j, \bar j;m_{s},m_{t})   \hat \cZ^{-1}  (  m _{s},m _{t};j', \bar j' ) 
&=& 
\delta_{j,j'}\delta_{\bar j, \bar j'}~.
   \eeqn 
 
   The columns and rows of  the matrix  $\hat \cZ$ and $\hat \cZ^{-1}  $  are labelled by  the  following pairs \footnote{If we furthermore have $p_1\ge \{p_2, p_3\} \ge p_4$, then the conditions for $\bm\Lambda_m$ are equivalent to $m_{24},m_{34}\ge 0,m_{24}+m_{34}\le L-1$.   
  }  
           \beqn\label{Ldm}
   \bm \Lambda_m&=&\Big\{(m_s,m_t)\Big| m_{ij}\in \bN,   \text{for } m_{ij}  \text{ given in } \eqref{m1234} \Big\}
  \\&=&
   \Big\{(m_s,m_t)\Big|   m_{s}  \in |\gamma|+2\bZ_{\ge 0} , m_{t}  \in \frac{ |\alpha-\beta|}{2}+2\bZ_{\ge 0} ,
 \nonumber   \\& & \qquad\qquad\quad
    m_{s} +m_t+2-\Sigma = -m_u\in - \frac{ |\alpha+\beta  |}{2} +2\bZ_{\le 0} \Big\}~,\qquad\qquad
 \label{Ldm22}
    \\ \label{Ldj}
  \bm     \Lambda_j&=&
   \Big    \{(j,\bar j)\Big| \frac12 \max(|\alpha|,| \beta|)\le  j\le \bar j \le \frac12 \max(|\alpha|,| \beta|)+  L-1  \Big\}~.
    \eeqn

   Now, we can invert Eq.~\eqref{Cstaulm} and obtain the following formula
 \be
    \sfC_s({\tau, \ell; j, \bar j } )
 =\sum_{(m_s,m_t)\in \bm\Lambda_m}
     \cC_{s}(\tau ,\ell;m_s,m_t)\hat   \cZ^{-1} ( m_{s},m_t ; {j,\bar j } ) ~.
 \ee
 This gives the explicit way of translating between internal Mellin space and internal spin space.

 \subsection{CFT data in  internal spin space}
 We now use the formalism above to translate the CFT data from internal Mellin space to internal spin space. 
 \subsubsection*{Case 1: $L=1$} 
 We start with the simplest case   $L=1$. For simplicity, we first assume $p_1\ge \{p_2, p_3\} \ge p_4$, so that $m_{24}=m_{34}=0$ while $m_s=\gamma,m_t=-\frac{\alpha }{2}-\frac{\beta }{2}-\gamma +\Sigma +2$. 
 The exchange operator has spin $j=\bar j =\frac12  \max(|p_ {34}|,|p_ {12}|)=\frac12(p_3+p_4)-1$, where we used the relation $L =\frac{p_4+p_3}{2}-\frac12 \max(p_ {34},p_ {12})=1$ in this case.  Using \eqref{zhatjms}, it is easy to compute
 \be
\hat\cZ=  \Gamma (\gamma +1) \Gamma \left(\frac{1}{2} (\alpha +\beta +2)\right) \Gamma \left(\frac{1}{2} (-\alpha -\gamma +\Sigma )\right) \Gamma \left(\frac{1}{2} (-\beta -\gamma +\Sigma )\right)~.
\ee
In particular,  for both $\EV{pp11}$ and  $\EV{p1p1}$, $\hat\cZ=  \Gamma(p)$.
 
We can also use formula \eqref{tau2eq} to compute the anomalous dimension
  \be
 \tau_2=\frac{6 \delta ^2-2 \delta +( {p_3}+ {p_4}-1)^2}{4 \sqrt{\delta }}~,
  \ee
which can be nicely rewritten as
   \be \label{tau2jjb}
 \tau_2 (\delta, \ell=2\delta; j, \bar j=j) =\frac{6 \delta ^2-2 \delta +( 2j+1)^2}{4 \sqrt{\delta }}~,
 \ee
 where the  quantum numbers indicate that  this multiplet sits on the leading Regge trajectory with equal internal spins.  As we will see later, this is a general feature.

Another simple example is given by   $\EV{4132}$, which  does not satisfy $p_1\ge \{p_2, p_3\} \ge p_4$, but we still have  $L=1$. In this case $m_s=0,m_t=1, j=\bar j=\frac32$ following from the formula \eqref{Ldm},\eqref{Ldj}. One can also compute $\hat\cZ=2$ and $\tau_2= {(6 \delta ^2-2 \delta +16)}/{(4 \sqrt{\delta })}$.
  
  Actually, the case of $L=1$ can be analyzed in full generality, thanks to the fact that 
  there is only one element in \eqref{Ldm} and \eqref{Ldj}.
From \eqref{Ldm22}, it is easy to  see that 
  \be\label{mstval}
  m_s=|\gamma|~, \qquad m_t=\frac{|\alpha-\beta|}{2}~, 
  \qquad m_u=\frac{|\alpha+\beta|}{2}~, \qquad m_s+m_t+m_u=\Sigma-2~,
  \ee
  just because there is only one element in $\bm \Lambda_m$. 
  
Meanwhile, the internal spins of the exchanged operator are given by
\be
  j=\bar j =\frac12  \max(|p_ {34}|,|p_ {12}|)
 =\frac12  \max(|\alpha |,| \beta |)
 =\frac{m_t+m_u}{2}
 =\frac{\Sigma-|\gamma|-2}{2}~,
\ee
where we used \eqref{mstval}.
Substituting into \eqref{tau2eq}, we can  compute the anomalous dimension
 \beqn
\tau_2 &=&\frac{\gamma ^2+6 \delta ^2-2 \delta -2  {m_s} (\Sigma -1)+\Sigma ^2-2 \Sigma +1}{4 \sqrt{\delta }}
\\&=&
\frac{6 \delta ^2-2 \delta +(\Sigma-|\gamma|-1)^2}{4 \sqrt{\delta }}
\\&=&
\frac{6 \delta ^2-2 \delta +( 2j+1)^2}{4 \sqrt{\delta }}~.
\eeqn
This is the same as eq. \eqref{tau2jjb}.

Using similar procedure, from \eqref{tau2subsubRg} we find that the anomalous dimension on the Regge trajectory $\ell=2\delta-2$ is:
\footnote{We have checked that up to $\delta=9$, this formula  reproduces the anomalous dimension computed from the $S$ and $T$ channel in the case $\EV{pp11}$  \cite{Jiang:2025oar}, where $j=0, (p-1)/2$, respectively. }
\be \label{tau2jjbsb}
\tau_2(\delta, 2\delta-2)=
\frac{6 \delta ^2-\frac{2 \delta }{3}+\frac{3020 \delta -504}{\delta  (\delta +24)-4}+(2 j+1)^2-124}{4 \sqrt{\delta }}~.
\ee

We can also compute $\hat \cZ$ using \eqref{zhatjms} which is a number in this case, 
\be
\hat \cZ=
\Gamma_S(m_s,m_t)=
\Gamma ( {|\gamma|}{ }+1 )\Gamma\Big(\frac{|\alpha-\beta|}{2}+1\Big)\Gamma\Big(\frac{|\alpha+\beta|}{2}+1\Big)
 ~,
\ee
where we used the formula   \eqref{GmS}.
Then all the OPE coefficients and amplitudes should be multiplied by a factor of $\cZ^{-1}$ when written as a function of the internal SU(2) spins. 
\footnote{This particularly implies that the Mellin amplitude in the supergravity limit for $\EV{pp11}$ should be given by \eqref{sugraM2} multiplied by a factor of $\hat \cZ^{-1}$, namely
$M_\text{SG}(s,t,\rho,\sigma)=- \frac{1}{\Gamma(p)}\Big(\frac{1}{s}-\frac{1}{t+1-p}-\frac{1}{\tilde u+1-p} \Big)$.
}
So the internal Mellin space and spin space essentially differ by an overall factor $\hat\cZ$.

    \subsubsection*{Case 2: $L>1$}
    Next we consider the case with $L>1$. This is more complicated because the exchanged multiplet can have different internal spins.  
    
  Let's illustrate with the simplest example $p_i=2$. Following \eqref{Ldm}  \eqref{Ldj}, it is easy to find that the possibles values of $(j,\bar j)$ and $(m_s,m_t)$ are given by
  \be\label{ldj}
  \bm \Lambda_j= 
  \Big\{(0,0),(0,1),(1,1) \Big\}~,
    \ee
    \be\label{ldm}
  \bm \Lambda_m= 
  \Big\{(0,0),(0,2),(2,0) \Big\}~.
    \ee
Further applying the formula \eqref{zhatjms}, one can compute all the values of    $\hat \cZ$  explicitly which are organized in the following matrix
  \be
\bm   {  \widehat    {\cZ}}=\left(
\begin{array}{ccc}
 0 & 0 & 2 \\
 -1 & 1 & 0 \\
 1 & 1 & -\frac{1}{2} \\
\end{array}
\right)~,\qquad
\qquad
\bm   {  \widehat    {\cZ}}^{-1}=
\left(
\begin{array}{ccc}
 \frac{1}{8} & -\frac{1}{2} & \frac{1}{2} \\
 \frac{1}{8} & \frac{1}{2} & \frac{1}{2} \\
 \frac{1}{2} & 0 & 0 \\
\end{array}
\right)~,
  \ee
  where we have also computed the inverse. Note the labeling for the columns and rows for $\hat\cZ$ and $\hat \cZ^{-1}$ are exchanged. 
  
  Now let us consider the function $F(m_s,m_t)=\sfa+\sfb m_s+\sfc m_t$, which is the general structure of the functions appearing in Mellin space. Evaluating the function at \eqref{ldm} gives the following vector
  \be
  \bm F=\Big(  \sfa,\sfa+2\sfc,\sfa+2\sfb\Big)~.
  \ee
  Then we can consider
  \be\label{Fzinv}
  \bm F \bm   {  \widehat    {\cZ}}^{-1}=\Big( \frac{3 \sfa+4\sfb+\sfc}{ 2} ,2\sfc,2\sfa+2\sfc \Big)~,
  \ee
where each entry corresponds to different spins $(j,\bar j)$ according to \eqref{ldj}.
These coefficients are the crucial ingredients when translating the CFT data back to internal spin space.  

Let us start with the leading OPE coefficients $f_0( \delta, \ell,m_s,m_t)=f_0(\delta, \ell)$ which is actually independent of $m_s,m_t$ \eqref{Asdel}.  This means that we are considering function $F=\sfa=f_0$ whose $\sfb=\sfc=0$. To go to spin space, we just need to incorporate the coefficients of $\sfa$ in \eqref{Fzinv}. As a result, 
$ \check f_0(\delta, \ell,j,\bar j)$ is given by
\footnote{We add the $\;\check {} \;$ symbol on top of $f,\tau$  etc to emphasize that these quantities are now functions of internal spins. }
\be
\check f_0(\delta, \ell,0,0 )=\frac32 f_0(\delta, \ell)~, \qquad
\check f_0(\delta, \ell,0,1 )=0~, \qquad
\check f_0(\delta, \ell,1 ,1 )=2 f_0(\delta, \ell)~,
\ee
where  $\check f_0(\delta, \ell,0,1 )=0$ implies that such a long multiplet with unequal spins $j\neq \bar j$ does not exchange in this case.  Identical discussions apply to higher $f_1,f_2 $, except that we also need to take into account the coefficients of $\sfb,\sfc$ because  $f_2$ does   depend on $m_s,m_t$ now. 

Next let us look at the anomalous dimension $\tau_2$.~\footnote{Note that $\tau_0,\tau_1$ are independent of internal spins. This is consistent with the fact that $\tau_0=2\sqrt{ \delta}$ which is obtained from matching the positions of the poles.}
A crucial thing is that in the block expansion \eqref{R1exp}, we   schematically have $f_2+\tau_2 f_0$, where $ \tau_2$ and $f_0$ appear in the product form. Both $f_0,f_2$ can be translated back to internal spin space following the prescription above. The translation is non-trivial even for $f_0$, although it is a constant function in Mellin space. Consequently, to translate $\tau_2$ back to the internal spin space, we should use an effective value of $m_s$, defined as the ratio of the coefficients of $\sfb$ and $\sfa$ in \eqref{Fzinv}.
 Explicitly, one gets  
  \be
 \bm m_s^\text{eff}=(\frac43, \quad , 0)~,
  \ee
  where the second entry  is empty, indicating  that the effective $m_s$ is not defined in this case.~\footnote{Similarly, one can also compute effective $m_t$, which reads
  $ \bm m_t^\text{eff}=(\frac13, \infty , 1)
$. The second entry is infinity, which may cause subtle effects, because $f_0=0$ for $j=0, \bar j=1$. 
 This subtlity is absent here as our $\tau_2$  in \eqref{mstval} does not depend on $m_t$!
  }
  This prescripition is also consistent with the previous case $L=1$, where $m_s^\text{eff}$ is just the $m_s$, which is also the ratio of the coefficients of $\sfb$ and $\sfa$ in Mellin space.
  
  Our previous results  \eqref{tau2eq} suggest that on the leading Regge trajectory 
  \be 
\tau_2(\delta, \ell=2\delta,m_s,m_t)=  \frac{6 \delta ^2-2 \delta -6  {m_s}+9}{4 \sqrt{\delta }}~.
  \ee
  In spin space,   $  \check\tau_2( \delta, \ell=2\delta, j,\bar j)$ is essentially given by $\tau_2$ above with $m_s$ replaced by $m_s^\text{eff}$
  \beqn
\check \tau_2(\delta, \ell=2\delta,0,0)&=&\frac{6 \delta ^2-2 \delta -6  \times\frac43+9}{4 \sqrt{\delta }}
 =\frac{6 \delta ^2-2 \delta +1}{4 \sqrt{\delta }}~, \qquad
\\
\check \tau_2(\delta, \ell=2\delta,1,1)&=&\frac{6 \delta ^2-2 \delta -6  \times 0+9}{4 \sqrt{\delta }}
 =\frac{6 \delta ^2-2 \delta +9}{4 \sqrt{\delta }}~, \qquad
  \eeqn
  It is easy to see that this nicely fits into the formula \eqref{tau2jjb}. This provides a strong consistency check of the prescription.   
 
The discussions above can be straightforwardly   generalized to all the general case of $\EV{p_1p_2p_3p_4}$. By going through many examples, we find that a linear function of the form   $F(m_s,m_t)=\sfa+\sfb m_s$ in Mellin space is always mapped to a function in spin space with support at equal spins, namely $F(j,\bar j)=0$ for $j\neq \bar j$. Looking at the CFT data on the leading Regge trajectory, it is easy to see that all the functions   indeed take  this form linear in $m_s$ and independent of $m_t$. This means that the exchanged long multiplets on the leading Regge trajectory always 
 have equal spins under the $SU(2)_L\times SU(2)_R$ R-symmetry.  
 This phenomenon  actually persists to all exchange long multiplet with even spacetime spin $\ell\in 2\bZ$, whose $f_0 \neq 0$ and $f_2$ is independent of $m_t$; and consequently only those with equal internal spins are exchanged. This seems to be a universal feature   whose  underling physical   reason  is worth exploring further. 
   The anomalous dimension for such equal-spin long multiplets on the leading Regge trajectory  is always given by  \eqref{tau2jjb}.

Let us also mention the Regge trajectory with $\ell=2\delta-1$. In this case we have  $f_0=f_1=0$, but $f_2\neq 0$ and is  linear in $m_t$  \eqref{f2subRg}, while $\tau_2$ is not determined because our  $\tau_2$ appears together with $f_0$ in the form $f_0\tau_2$, which is zero due to $f_0=0$.   We can similarly use the previous matrix $\hat\cZ$ to translate $f_2$ back to internal spin space. 
 
 Finally, we use our prescription to derive a simple class of three point function in D1--D5 CFT, arising from two identical half-BPS tensor operators and one long operator on the leading Regge trajectory. 
 This can be derived by considering the four point function   $\EV{ppqq}$.  Using our procedure above, for a function  
  $F(m_s,m_t)=\sfa+\sfb m_s$ in internal Mellin space, one can compute the corresponding function in the internal spin space:~ \footnote{We have verified this formula  for  $p,q\le 12$.} 
    \be\label{Fjjbar}
  \check F(j ,  j )=
\frac{(p+q-1)!\,(2j+1)\,((2j)!)^2}{(j!)^4\,(p-1-j)!\,(p+j)!\,(q-1-j)!\,(q+j)!}
\Big( \sfa+\frac{ (p-1) p+(q-1) q-2 j (j+1)}{p+q-1} \sfb\Big) 
 \ee
if $ 0\le j\le \min\{p-1,q-1\} $ and zero otherwise. 
 
 \bigskip
 
%
%
 
 Following previous formula \eqref{fexpres}, the  OPE coefficient  product  in Mellin space is given by
   \beqn\label{fexpres222}
 \cC_s(\tau, \ell=2\delta; m_s,m_t)  &=&
\frac 
{\lambda ^{\frac{\Sigma -1}{2}+\frac{3}{4}} (1+\delta_{\ell,0})}
 {4^{\ell+\tau }  \sin  \frac{   \pi  (\alpha -\tau )}{2}  \sin  \frac{   \pi  (\beta -\tau )}{2}    \Gamma (\Sigma )  i^{-\alpha -\beta +2 \Sigma } }
 f(\tau, \ell; m_s,m_t) 
\\&=&
\frac 
{  (-1)^{ p+q }  \lambda ^{\frac{p+q }{2}+\frac{1}{4}} }
 {4^{2\delta+\tau }  \sin^2  \frac{   \pi    \tau  }{2}     \Gamma (p+q )  }
( f_0 + f_1 \lambda^{-1/4} + f_2 \lambda^{-1/2} + f_3 \lambda^{-3/4} + \cdots)~,
\qquad\qquad
\eeqn
where all $f_i$'s have been computed in section~\ref{AdSVS}.
Translating to spin space yields: 
\footnote{As described above, for unequal spins $j\neq \bar j$, one has $ \check F=0$, so such long multiplet does not exchange between half-BPS tensors.  }
   \beqn\label{fexpres23}
 \sfC_s(\tau, \ell=2\delta; j ,\bar j=j)  &=&
\frac 
{  (-1)^{ p+q }  \lambda ^{\frac{p+q }{2}+\frac{1}{4}} }
 {4^{2\delta+\tau }  \sin^2  \frac{   \pi    \tau  }{2}     \Gamma (p+q )  }
( \check f_0 + \check  f_1 \lambda^{-1/4} + \check  f_2 \lambda^{-1/2} + \check  f_3 \lambda^{-3/4} + \cdots)
\\ &=&
\frac  {  (-1)^{ p+q }  \lambda ^{\frac{p+q }{2}+\frac{1}{4}} }
 {4^{2\delta+\tau }  \sin^2  \frac{   \pi    \tau  }{2}     \Gamma (p+q )  }\check f_0 
\Big(1 + \check  f_1/ \check f_0  \lambda^{-1/4} + \check  f_2/ \check f_0  \lambda^{-1/2} + \check  f_3/ \check f_0  \lambda^{-3/4} + \cdots \Big)~.
\nonumber
\\
\eeqn
Note that essentially, we have
\be
 \check  f_i/ \check  f_0=f_i/f_0|_{m_s\to m_s^\text{eff}=\frac{ (p-1) p+(q-1) q-2 j (j+1)}{p+q-1}}~,
 \ee
  and 
\be
\check f_0=\frac{\pi ^3 2^{-4 \delta -1} \delta ^{2 \delta +\Sigma -\frac{3}{2}}}{\Gamma (\delta )^2}
\frac{(p+q-1)!\,(2j+1)\,((2j)!)^2}{(j!)^4\,(p-1-j)!\,(p+j)!\,(q-1-j)!\,(q+j)!} ~.
\ee
where we used \eqref{f0evenspin} and \eqref{Fjjbar}.
 
Performing the computation explicitly, one finds that  $\sfC_s$ takes the factorized form:
\be
 \sfC_s(\tau, \ell=2\delta; j ,\bar j=j) =\sfC _{pp\tau}\sfC_{qq\tau}~,
\ee
which   $\sfC_{pp\tau}$ is exactly the   OPE coefficient or three-point function between two half-BPS tensor operator  of dimension $p$ and one long-multiplet on the leading Regge trajectory, which has conformal dimension $\tau+\ell$, spacetime spin $\ell=2\delta$ and internal spins $\bar j=j$.  Note that the factorization structure is  a non-trivial consistent check; for example, the factor $\Gamma(\Sigma)=\Gamma(p+q)$ \eqref{fexpres222}, which is an  obstruction  for the factorization,  is exactly cancelled by the same factor $(p+q-1)!$ in  \eqref{Fjjbar}.

Explicitly, the three-point function reads
 \begin{align}\label{pptauOPE}
\sfC_{pp\tau}=&
\frac{\pi ^{3/2} \sqrt{2 j+1} (-1)^p 2^{-4 \delta -\tau -\frac{1}{2}} \Gamma (2 j+1) \delta ^{\delta +p-\frac{3}{4}} \lambda ^{\frac{p}{2}+\frac{1}{8}}  }{\Gamma (\delta ) \Gamma (j+1)^2 \Gamma (p-j) \Gamma (j+p+1)\sin \left(\frac{\pi  \tau }{2}\right)}
  \Bigg\{
  1
 + \lambda^{-\frac14} \left[-\frac{5}{8\sqrt{\delta}}-\sqrt{\delta}\right]
\nonumber \\  
&+\lambda^{-\frac12}
\Bigg[
\frac{11}{24}+j+j^{2}-\frac{p}{2}-p^{2}
+\frac{ \frac{17}{128}+\frac{p}{3}-\frac{p^{3}}{3}+j\!\left(\frac14+p\right)+j^{2}\!\left(\frac14+p\right)}{\delta}
\nonumber \\ &
\qquad \qquad 
+\frac18(9+4p)\,\delta
-\frac{7}{12}\delta^{2}
+\delta^{2}\zeta(3)
\Bigg]
\nonumber\\ 
&+
\lambda^{-\frac34} \Bigg[
\frac{123-640p+640p^{3}-480j(-1+4p)-480j^{2}(-1+4p)}{3072\,\delta^{3/2}}
\nonumber\\[6pt]
&\qquad \qquad 
+\frac{-119-8p+240p^{2}+128p^{3}-48j(3+8p)-48j^{2}(3+8p)}{384\,\sqrt{\delta}}
\nonumber \\ 
& \qquad \qquad 
+\left(-\frac{181}{192}-j-j^{2}+\frac{3p}{16}+p^{2}\right)\sqrt{\delta}
+\left(-\frac{1}{96}-\frac{p}{2}\right)\delta^{3/2}
\nonumber\\ 
&\qquad \qquad 
+\frac{7}{12}\delta^{5/2}
+\left(-\frac{5}{8}\delta^{3/2}-\delta^{5/2}\right)\zeta(3)
\Bigg]
+\mathcal O(\lambda^{-1})
   \Bigg\}~,
\end{align} 
while the twist and spacetime spins of the long operator are
 \be\label{tau3full}
 \tau\equiv\Delta-\ell=2\sqrt{ \delta} \lambda^{\frac14}-2\delta-1+\frac{6 \delta ^2-2 \delta +(2j+1)^2}{4 \sqrt{\delta } \lambda^{\frac14}}+\cO(\lambda^{-3/4}) ~, \qquad
 \ell=2\delta~, \qquad \delta\in \bZ_{>0}~.
 \ee
 
 It is straightforward to generalize and find the OPE coefficients involving two different half-BPS operators, namely $\sfC_{pp'\tau}$, although finding an explicit formula valid for general $p,p'$ is non-trivial. These results give the explicit analytic CFT data in D1--D5 CFT.

  
\acknowledgments

 We   thank Deliang Zhong   for related discussions. 
This work  was supported by the startup grant at SIMIS and   the Shanghai Pujiang Program (No. 25PJA128). 
 
\appendix

\section{Single-valued multiple polylogarithms} \label{apd-SVMPL}

The multiple polylogarithms (MPLs) of weight $r$ are defined though iterated integrals  \cite{Goncharov:2001iea}
\be
L_{a_1 a_2 \ldots a_r}(z) \coloneq \int_0^z \frac{\dd t}{t-a_1} L_{a_2 \ldots a_r}(t)~,
\ee
where the words $a_i$ take values in $\{0,1\}$. Explicitly, the multiple polylogarithms up to weight three are given in terms of combinations of logarithm, polylogarithms and zeta values as follows:
\be
\begin{aligned}
L_{0^p}(z) =&\frac{\log ^p z}{p!}~, \\
 L_{1^p}(z) =&\frac{\log ^p(1-z)}{p!}~, \\
L_{01}(z)= & -\operatorname{Li}_2(z)~, \\
 L_{10}(z)=&\operatorname{Li}_2(z)+\log (1-z) \log (z)~, \\
  L_{010}(z)=&2 \operatorname{Li}_3(z)-\operatorname{Li}_2(z) \log (z)~,\\
L_{001}(z)= & -\operatorname{Li}_3(z)~, \\
 L_{100}(z)= & -\operatorname{Li}_3(z)+\operatorname{Li}_2(z) \log (z)+\frac{1}{2} \log (1-z) \log ^2(z)~,\\
L_{101}(z)= & 2 \operatorname{Li}_3(1-z)-2 \zeta(3)-\log (1-z)\left(2 \operatorname{Li}_2(1-z)\right.   \left.+\operatorname{Li}_2(z)+\log (1-z) \log (z)\right) ~,\\
L_{110}(z)= & -\operatorname{Li}_3(1-z)+\frac{1}{6} \pi^2 \log (1-z)+\zeta(3)~, \\
L_{011}(z)= & -\operatorname{Li}_3(1-z)+\operatorname{Li}_2(1-z) \log (1-z)  +\frac{1}{2} \log (z) \log ^2(1-z)+\zeta(3)~,
\end{aligned}
\ee
where $\operatorname{Li}_n(z)$ denotes the polylogarithms of order $n$. 

MPLs are holomorphic functions of $z$ but they are not singe-valued. The single-valued version of MPLs can be constructed systematically. In particular, up to weight three, single-valued multiple polylogarithms (SVMPLs) $\mathcal{L}$ are constructed from MPLs as
\be
\begin{aligned}
& \mathcal{L}_{000}(z)=L_{000}(z)+L_{000}(\bar{z})+L_{00}(z) L_0(\bar{z})+L_0(z) L_{00}(\bar{z}), \\
& \mathcal{L}_{001}(z)=L_{001}(z)+L_{100}(\bar{z})+L_{00}(z) L_1(\bar{z})+L_0(z) L_{10}(\bar{z}), \\
& \mathcal{L}_{010}(z)=L_{010}(z)+L_{010}(\bar{z})+L_{01}(z) L_0(\bar{z})+L_0(z) L_{01}(\bar{z}), \\
& \mathcal{L}_{100}(z)=L_{100}(z)+L_{001}(\bar{z})+L_{10}(z) L_0(\bar{z})+L_1(z) L_{00}(\bar{z}), \\
& \mathcal{L}_{110}(z)=L_{110}(z)+L_{011}(\bar{z})+L_{11}(z) L_0(\bar{z})+L_1(z) L_{01}(\bar{z}), \\
& \mathcal{L}_{101}(z)=L_{101}(z)+L_{101}(\bar{z})+L_{10}(z) L_1(\bar{z})+L_1(z) L_{10}(\bar{z}), \\
& \mathcal{L}_{011}(z)=L_{011}(z)+L_{110}(\bar{z})+L_{01}(z) L_1(\bar{z})+L_0(z) L_{11}(\bar{z}), \\
& \mathcal{L}_{111}(z)=L_{111}(z)+L_{111}(\bar{z})+L_{11}(z) L_1(\bar{z})+L_1(z) L_{11}(\bar{z}) .
\end{aligned}
\ee

MPLs and SVMPLs can be implemented using package $\mathtt{PolyLogTools}$ \cite{Duhr:2019tlz,Maitre:2005uu}. 

\section{Mack Polynomials} \label{apd-mackpol}

As shown in   \cite{Jiang:2025oar}, which follows from  \cite{Dolan:2011dv},  the Mack polynomial $Q^{\Delta_{12}, \Delta_{34}, \tau}_{\ell, m}(s)$ admits the following double sum form:
\beqn
Q^{\Delta_{12}, \Delta_{34}, \tau}_{\ell, m}(s) &= 
 \sum\limits_{k=0}^\ell \sum\limits_{n=0}^{\ell - k}
(-m)_k \left( m + \frac{s + \tau}{2} \right)_n 
\mu(\ell, k, n, \Delta_{12}, \Delta_{34}, \tau)~,
\eeqn
where $(a)_b$ denotes the Pochhammer symbol, and $\mu$ is given by
\begin{align}
\mu(\ell, k, n, \Delta_{12}, \Delta_{34}, \tau) &=
(-1)^{k+n+\ell } 
\frac{2^{ \ell} \, \ell! \, \Gamma(\ell + \tau - 1)}{\Gamma(2\ell + \tau - 1)}
\frac{ (\ell + \tau - 1)_n}{k! \, n! \, (-k - n + \ell)!}
\left( n + \frac{d} {2}-1+ \frac{\Delta_{34} - \Delta_{12}}{2} \right)_k \notag \\
&\quad \times 
\left( k + n - \frac{\Delta_{12}}{2} + \frac{\tau}{2} \right)_{-k - n + \ell}
\left( k + n + \frac{\Delta_{34}}{2} + \frac{\tau}{2} \right)_{-k - n + \ell} \notag \\
&\quad \times 
{}_4F_3\left(
\begin{array}{c}
-k, \, 3-d - n - \ell, \, 1-\frac{d}{2} + \frac{\Delta_{12}}{2} + \frac{\tau}{2}, \, 1-\frac{d}{2} - \frac{\Delta_{34}}{2} + \frac{\tau}{2} \\
2-\frac{d}{2}-\ell, \,2-\frac{d}{2}+ \frac{\Delta_{12}}{2} - \frac{\Delta_{34}}{2} - k - n, \,  2-d + \tau
\end{array}
; 1
\right)\, .
\end{align}

 The normalization  of Mack polynomial is chosen such that $ Q_{\ell,m}^{\tau,d}(s)=s^\ell+\cdots$. For $m=0$, the Mack polynomial reduces to 
  \be
  Q_{\ell,0}^{\tau,d}(s)=  \frac{(-2)^\ell \left(\frac{\tau }{2}-\frac{\Delta_{12}}{2}\right)_\ell \left(\frac{\Delta_{34}}{2}+\frac{\tau }{2}\right)_\ell  }{(\ell+\tau -1)_\ell}\, _3F_2\left(-\ell,\frac{s}{2}+\frac{\tau }{2},\ell+\tau -1;\frac{\tau }{2}-\frac{\Delta_{12}}{2},\frac{\Delta_{34}}{2}+\frac{\tau }{2};1\right)~.\\
  \ee

The normalization factor $\kappa$ used in the main text reads 
\be\begin{aligned} 
\kappa_{\ell, m, \tau,d}^{\left(p_1, p_2, p_3, p_4\right)} & 
= \frac{-2^{1-\ell}(\ell+\tau-1)_{\ell} \Gamma(2 \ell+\tau)  }{m!(1-\frac{d}{2}+\ell+\tau)_m }
\frac{1}{ \Gamma\left(-m+\frac{p_1+p_2}{2} -\frac{\tau}{2}\right) \Gamma\left(-m+\frac{p_3+p_4}{2} -\frac{\tau}{2}\right)}
 \\ & \times 
 \frac{1}{ 
\Gamma\left(-\frac{p_1-p_2}{2}+\ell+\frac{\tau}{2}\right)
\Gamma\left(\frac{p_1-p_2}{2}+\ell+\frac{\tau}{2}\right)
 \Gamma\left(-\frac{p_3-p_4}{2}+\ell+\frac{\tau}{2}\right) \Gamma\left(\frac{p_3-p_4}{2}+\ell+\frac{\tau}{2}\right) 
}\, .
\end{aligned}
\ee 
In this paper, we consider two dimensional field theory, so should set $d$ to 2. 
\footnote{It is worth mentioning  that to get Mellin block in two dimensions, one need to consider general $d$ first and then take the limit $d\to 2$.
 }

\section{Transformation function in internal spaces}\label{MackS}

Our goal in this appendix is to derive the equation \eqref{zhatjmeq}, which is crucial in bridging the    internal spin space and internal Mellin space. 

We start with some essential identities. 
For $m-n\in \bZ$, we have the following identity 
\be
y^m \bar y^n+y^n \bar y^m=2 \sigma ^{-\left(\frac{m}{2}+\frac{n}{2}\right)} T_{n-m}\left(\frac{-\rho +\sigma +1}{2 \sqrt{\sigma }}\right)~,
\ee
where $T_l$ is the Chebyshev polynomial. 
 This   allows us to  further    rewrite it as
 \be\label{ybtheta}
y^m \bar y^n+y^n \bar y^m
=2\sum_{p=-\max(m,n)}^{-\min(m,n)}\sum_{q=0}^{|m-n|}\Theta_{m,n}({p,q}) \sigma^p \rho^q~,
\ee
where the coefficient is given by 
\be \label{thetamnpq}
\Theta_{m,n}({p,q}) 
=
 \sum_{l=0}^{\lfloor k/2\rfloor} 
\frac{(-1)^{q+l}\;k\;\Gamma(k-l )}
{2 \Gamma(l+1)\,\Gamma(q+1)\,\Gamma(p+\max(m,n)-l+1)\,\Gamma( -l-q-p-\min(m,n)+1)}~,
\ee
for    $k\equiv |m-n|>0$, and $\Theta_{m,m}({p,q})  =  \delta_{m+p}\delta_{q,0}$  for $k=0$.

 The other useful identity   is
 \be  
 {}_2F_1(-m,-n;\,-r;\,x)
=
\sum_{k=0}^{\min(m,n)}
\frac{m!\,n!}{(m-k)!\,(n-k)!}\,
\frac{1}{(-r)_k}\,
\frac{x^k}{k!}~,
\qquad \bigl(0\le\min(m,n)\le r\bigr)~,
 \ee
 where $m,n\in  \bN$. Note that $(y)_k=\Gamma(y+k)/\Gamma(y)=(-1)^k (-k-y+1)_n=(-1)^k\Gamma(1-y)/\Gamma(1-y-k)$. This immediately gives 
 \be\label{kappaf21}
\kappa_{-2j}^{(2r ,2s)}(x)= x^{-j}{}_2F_1(-j+r,-j+s;\,-2j;\,x)
 =\sum_{n=\max(r,s)}^j
 \frac{(j-r)!\,(j-s)!}{(n-r)!\,(n-s)!}\,
 \frac{ (j+n  )! (-1)^ {j-n}}{(j-n)!(2j)!}x^{-n}~,
 \ee
 for $j-r,j-s\in \bN$.
 Note that in the above formula, we can just sum $n$ over all integers/half-integers ($n\in \bZ+r$), because the factorials in the denominator impose the conditions
 $n \ge r, n\ge s, j \ge n$.
 
 The identity \eqref{kappaf21} enables us to write the R-symmetry conformal block    Eq.~\eqref{Rblock2} explicitly as
 \beqn
 \hat Z_{ j, \bar j}^{(2r ,2s)}(\rho, \sigma)
& =&\frac12
{ \sigma} ^{-\frac{p_3+p _4 }{2}+L} \sum_{n,\bar n  }
 \frac{(j-r)!\,(j-s)!}{(n-r)!\,(n-s)!}\,  
\frac{ (\bar j-r)!\,(\bar  j-s)!}{(\bar n-r)!\,(\bar n-s)!}\, 
  \\&&\qquad\qquad\qquad \times
 \frac{ (j+n  )! (-1)^{j-n}}{(j-n)!(2j)!}
 \frac{ (\bar j+\bar n  )! (-1)^{\bar j-\bar n}}{(\bar j-\bar n)!(2\bar j)!}
   ( y ^{-\bar n} \bar y^{-n}+y ^{-n} \bar y^{-\bar n})
   \\
   & =&
{ \sigma} ^{-\frac{p_3+p _4 }{2}+L} \sum_{n,\bar n  ,p,q}
 \frac{(j-r)!\,(j-s)!}{(n-r)!\,(n-s)!}\,  
\frac{ (\bar j-r)!\,(\bar  j-s)!}{(\bar n-r)!\,(\bar n-s)!}\, 
 \\&&\qquad\qquad\qquad \times
 \frac{ (j+n  )! (-1)^{j-n}}{(j-n)!(2j)!}
 \frac{ (\bar j+\bar n  )! (-1)^{\bar j-\bar n}}{(\bar j-\bar n)!(2\bar j)!}
\Theta_{-n, -\bar n}(p,q) \sigma^p \rho^q~, 
   \qquad\qquad
 \eeqn
 where $2r=p_{12} \ge 0,2s=p_{34} \ge 0$ and we used the identity Eq.~\eqref{ybtheta}. This equation should be compared with  \eqref{Zseq232}:
  \be 
 \hat Z_{j,\bar j}^{( p_{12}, p_{34})}(  \rho,\sigma)  
 =\sum_{m_{24},m_{34} }
\frac{   \rho^{m_{24}}\sigma^{L-1-m_{24}-m_{34}}  
}{  \Gamma_{S} (m_s,m_t)}
\hat \cZ ^{( p_{12}, p_{34})} (j,\bar j; m_{s},m_t)~
\ee
 yielding 

 \beqn\label{zhatjm}
\hat \cZ ^{(2r ,2s)} ( j, \bar j; m_{s},m_{t})&=& 
 \sum_{n,\bar n    }
 \frac{(j-r)!\,(j-s)!}{(n-r)!\,(n-s)!}\,   
   \frac{(\bar j-r)!\,(\bar j-s)!}{(\bar n-r)!\,(\bar n-s)!}\,  \frac{ (\bar j+\bar  n  )! (-1)^{j-n}}{(j-n)!(2j)!}
 \frac{ (\bar j+\bar n  )! (-1)^{\bar j-\bar n}}{(\bar j-\bar n)!(2\bar j)!}
\nonumber \\&&   \qquad 
\times        \Gamma_{S}(m_{s},m_{t})
\Theta_{\bar n,n}
\Big({\frac{2m_t+p_1-p_2+p_3-p_4}{4},\frac{-2-m_s-m_t+p_2+p_4}{2}}\Big)~.
   \nonumber\\
 \eeqn
  This equation hence builds the bridge for connecting  internal spin space   and internal Mellin space.  It can be regarded as the internal space analogue of the Mack polynomial in Appendix~\ref{apd-mackpol}.

\bibliographystyle{JHEP.bst} 
\bibliography{AdS3}

\end{document}